# Optical Properties of Nanocellulose and Nanocellulose-based Composites for Optoelectronic Applications

Jonas Rashidi [1,2,*]

[1] Institute of Materials, École Polytechnique Fédérale de Lausanne (EPFL), Station 12, Lausanne CH-1015, Switzerland

[2] Department of Advanced Physics and Chemistry, Farhangian University, Iran

**\*** Correspondence: Junassraschidie@gmail.com

**Abstract**

In this review, we first provided a brief overview of cellulose structures, categorizing various types of nanocellulose based on their production methods, and discussed different photonic structures such as cholesteric phases, photonic crystals, distributed Bragg reflectors, and inverse photonic structures. We then reviewed some models that predict optical properties, including refractive index, scattering, reflection, and birefringence, highlighting factors that can influence these properties, such as electric field, surface roughness, temperature, and density, among others. Following this, we explored recent advancements in predicting optical properties, particularly the use of AI-driven algorithms to enhance the prediction of optical behaviors in nanomaterials. Finally, we provided an in-depth discussion of the latest developments in nonlinear optical phenomena and chiroptical responses in nanocellulose-based materials, focusing on circularly polarized luminescence, structural colors, structural correlations, the impact of enantiomers, and more.

# Introduction

As the need to reduce humanity's dependence on polymer materials derived from fossil fuels grows increasingly urgent, efforts have been directed toward developing functional biocompatible materials from natural sources. Among these, cellulose has been widely studied due to its versatility and natural availability. Composed of β-D-glucopyranose units connected by glycosidic linkages, cellulose is a natural polymer that plays a crucial role in providing structural strength to plant cell walls and protecting bacterial cells (Figure 1). [1,2] Materials utilizing cellulose at different length scales benefit from its exceptional mechanical properties, such as Young's modulus, and thermal stability [3]. However, nanocellulose-based materials can be classified based on their manufacturing methods into four distinct categories, as proposed by Klemm et al.[4]: (i) Nanocellulose Whiskers (CW):[5] Also known as cellulose nanocrystals (CNCs), these are rod-like nanostructures obtained by acid hydrolysis of cellulose fibers. They possess a high aspect ratio, excellent mechanical strength, and liquid crystalline properties, (ii) Bacterial Nanocellulose (BC):[6] Produced by bacterial fermentation, particularly from species such as Komagataeibacter hansenii, BC features a highly pure and highly crystalline cellulose network with remarkable mechanical strength, biocompatibility, and water retention capacity, (iii) Micro/nanofibrillated Cellulose (MFC):[7] Obtained by mechanical shearing (high-pressure homogenization, grinding, or refining) of wood pulp, MFC consists of entangled cellulose nanofibrils with high surface area and enhanced flexibility, (iv) Electrospun Cellulose Nanofiber (CNF):[8] Fabricated through electrospinning, this type of nanocellulose results in continuous fibers with nanoscale diameters, providing tunable porosity and high specific surface area.

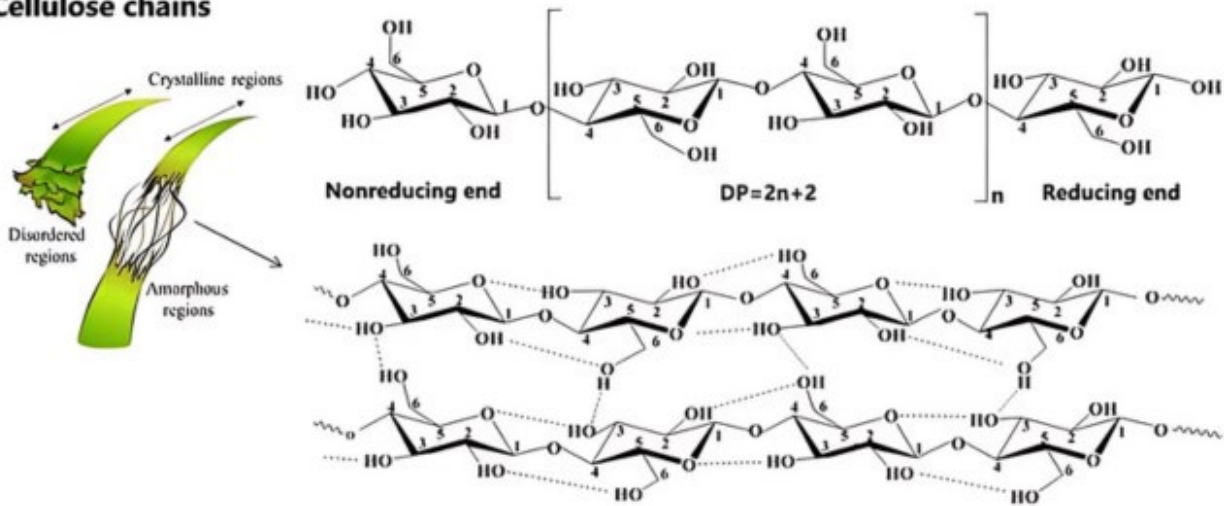

Figure 1. A cellobiose unit consists of two β-D-glucopyranose units linked by a β-1,4-glycosidic bond. Copyright 2021 the authors. Published by Springer under a Creative Commons CC BY License. Ref:1

In recent years, various innovative approaches have been utilized to develop novel nanocellulose-based materials. Among their most promising characteristics are their optical properties. The alterations in light properties that occur when it interacts with a medium are referred to as the optical properties of that medium. These optical properties are influenced by the medium's physical and chemical characteristics, such as surface roughness and dangling bonds. However, measuring optical properties is generally simpler and more straightforward than analyzing the physical and chemical properties and their complex correlations with optical behavior. Consequently, optical properties are often studied as a means to gain insights into other material properties. Some optical properties are more commonly observed, including scattering, refraction, dispersion, transmission, absorbance, and birefringence. [9] Birefringence, for instance, can arise when nanocellulose-based materials self-assemble into 2-fold twisting and left-handed chiral nematic phases [10], where the refractive index varies based on the polarization and propagation of incident electromagnetic radiation, resulting in the circular polarization of the incoming irradiation. They also exhibit tunable birefringence, making them highly suitable for a wide range of applications, including light modulators, and optical storage discs. [11] This property, combined with an uneven distribution of left- and right-handed chirality within the material, not only induces circular polarization but also forms ideal scaffolds for circularly polarized luminescence (CPL). [10,12]

Photonic Structures

To gain better control over photons and their interactions, including polaritons and exciton-photon coupling, photonic structures have been developed. These engineered materials regulate the propagation, emission, and interaction of light through periodic optical patterns or refractive index modulation. Each photonic material possesses a specific photonic band gap (PBG) that inhibits the propagation of certain wavelengths. Various photonic structures exist for cellulose-based materials, including photonic crystals (PC), Bragg stacks, and chiral nematic liquid crystals. Here, we briefly introduce their definitions.[13] The photonic structure can be classified as either ordered or disordered structures. Ordered structures include the cholesteric phase, photonic crystal, and distributed Bragg reflectors.[14] Liquid crystal is a state of matter that flows like a liquid while maintaining an ordered structure similar to a solid crystal.[15] Most liquid crystalline phases exhibit strong dipole interactions and possess a rigid, rod-like structure, leading to anisotropic behavior under specific conditions. Depending on the properties of the liquid crystal and external influences, various configurations can be expected (Figure 2). As mentioned below, chiral nematic phases exhibit high positional and orientational order while maintaining a helical axis with a defined direction.[16] For instance, Fu et al. synthesized oxalated cellulose nanocrystals, leading to the formation of two distinct phases: the nematic phase and the cholesteric phase, using the multilayer spin-coating technique. [17]

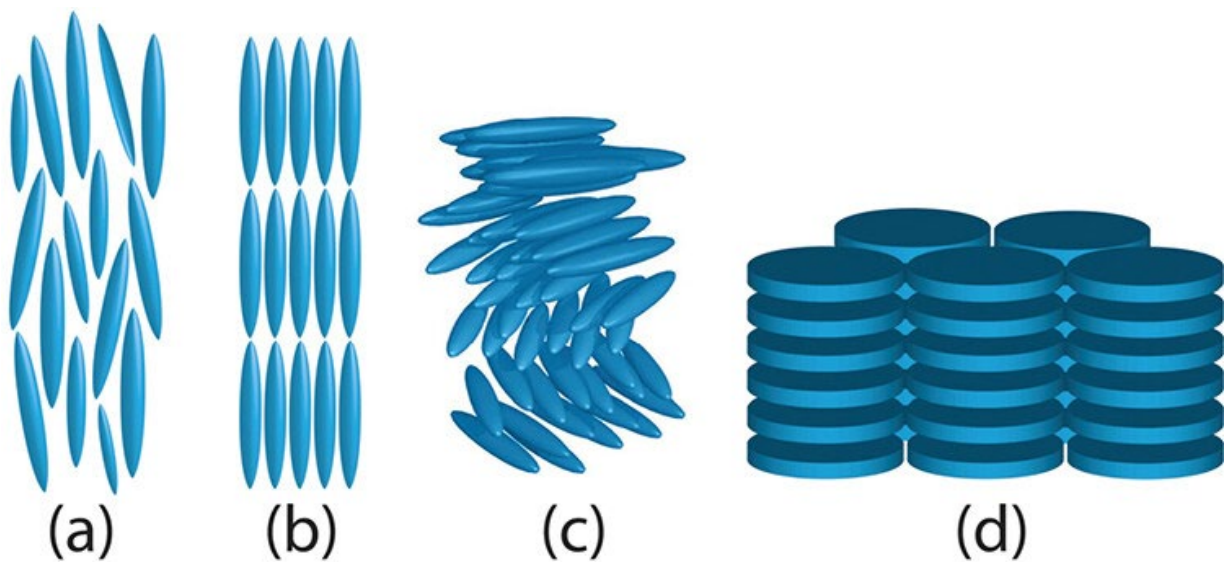

Figure 2. (a) The nematic phase exhibits high orientational order with low positional order. (b) The smectic phase is characterized by both high orientational and positional order. (c) The cholesteric phase maintains high orientational and positional order while incorporating a helical axis with a defined direction. (d) The columnar liquid crystalline phase forms a regular two-dimensional lattice where translational invariance is disrupted in two directions. There is another general phase though that remains unmentioned, the isotropic phase with no general order at all. Copyright 2024 the authors. Published by ACS under a Creative Commons CC BY License. [16]

A photonic crystal (PC) is a nanostructured material with a periodically varying refractive index. In other words, its dielectric function exhibits periodicity in one, two, or three spatial dimensions (1D, 2D, or 3D), analogous to the periodic potential in electronic crystal structures. [18]. In a study, Jia et al. developed a tunable photonic band gap, multi-stimuli-responsive cellulose nanocrystals-poly(ethylene glycol) acrylate-pentanediol composite to mimic chameleon skin. The ternary supramolecular film responded to humidity, solvents, wetting gases, and pressure. Under pressure, the film shrank and exhibited a blue-shift in color, with the ability to recover upon immersion in ethanol followed by slow evaporation. Additionally, exposure to wetting gas caused the film to curve to one side, which reversed upon drying.[19] A distributed Bragg reflector (DBR) consists of alternating layers with high and low refractive indices, where the layer thickness satisfies the "quarter-wave" condition at a specific wavelength. This photonic structure exploits successive reflections at dielectric interfaces. As the number of high/low refractive index layers increases, stronger constructive interference enhances reflectivity.[20] However, a major challenge lies in the requirement for materials with a higher refractive index contrast, while most polymeric materials typically exhibit refractive indices between 1.4 and 1.6.[21] In a study by Seike et al., combinations of cellulose acetate (CA) and copper (I) were used to fabricate DBRs consisting of 3 to 11 layers, exhibiting peak reflectance at 700 nm. They developed perovskite solar cells integrated with these DBRs using a coating-based fabrication technique. In the case of the 11-layer DBR, the structure enhanced light absorption in the perovskite cells but led to a 39.0% and 41.2% reduction in power conversion efficiency and short-circuit current density, respectively, compared to perovskite cells without DBRs.[22] Until now, we have discussed photonic structures composed of purely ordered materials. However, what if disorder is introduced in a controlled manner to achieve specific optical properties? Disordered photonic structures lack long-range periodicity, leading to a high degree of light scattering, which results in materials that appear whiter or more opaque. For example, using this strategy, the Lepidiota stigma beetle achieves an exceptionally white appearance with minimal material consumption to maximize multiple light scattering.[14] One important parameter in measuring scattering is the transport mean free path, which represents the average distance a photon or electron travels before its direction becomes randomized due to multiple scattering events. This parameter accounts for the angular distribution of scattering, making it crucial for understanding light transport in disordered photonic structures and thin films. For further details, refer to the recommended reference Leonetti et al.[23] Minimizing the transport mean free path maximizes scattering, resulting in whiter and more opaque materials. Researchers aim to reduce this parameter as much as possible. For instance, Caixeiro et al.[24] fabricated a disordered CNC structure with a transport mean free path of 3.5 μm, significantly lower than the 20 μm observed in ordinary microfiber-based paper. This structure exhibited four times stronger scattering than typical paper. The material was composed of CNC and colloidal monodisperse polystyrene (PS) spheres with a diameter of 1.27 μm, using a dry mass fraction of CNC/PS at 2/3. The selection of spheres with a diameter equal to half the wavelength was critical for maximizing scattering. Before etching the PS with toluene, controlled water evaporation resulted in an inverse photonic cellulose structure, as illustrated in Figure 3.

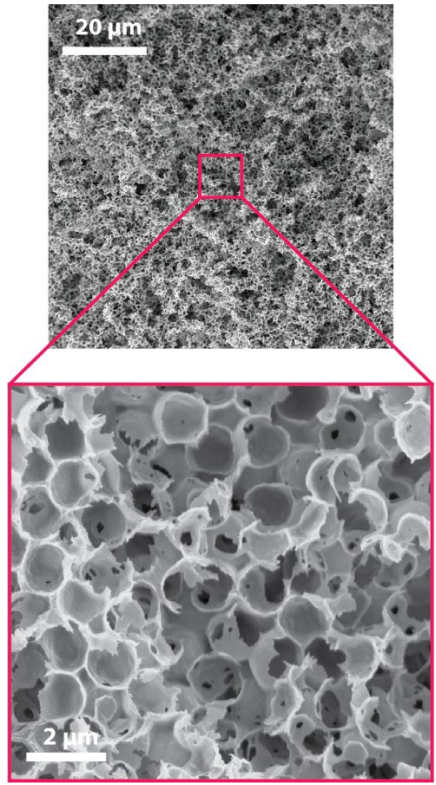

Figure 3. SEM images of a free-standing cellulose inverse photonic glass, displaying spherical voids with a diameter of 1.3 µm and smaller voids at points where PS spheres were in contact before etching. The structure exhibits homogeneous circular openings, characteristic of a close-packed arrangement across the sample.

While nearly all optical properties of cellulose are well-documented, a comprehensive review that addresses and categorizes the latest findings could act as a stepping stone to further advance ongoing research. In this context, we adopt both a retrospective and a prospective perspective to highlight some of the most significant optical properties of nanocellulose-based materials, including structural correlations, birefringence, intrinsic luminescence, nonlinear optical properties, optical bandgap ($E_g$), and their prevalent applications. Before beginning this section, we will provide a brief explanation of the primary optical properties, such as refractive indices, transmission, and reflection.

1) Modeling Optical Properties: Key Features and Approaches

By adopting a nonrelativistic microscopic perspective, the phase velocity of an electromagnetic wave in a medium is influenced by an external perturbation due to the interaction between the external electric field and the atomic charges. As electromagnetic fields oscillate, the charges within the material are displaced and excited to energy levels. These charges subsequently emit electromagnetic waves that can either synchronize with or deviate from the primary wave. This interaction leads to constructive or destructive interference, resulting in a superposition of waves that alters the phase velocity—either increasing or decreasing it—relative to the original light source, while typically maintaining the same frequency. Generally, the refractive index ($n$) is defined as the ratio of the speed of light in a vacuum without perturbation ($c$) to the speed of light in a given medium, which can be either the phase velocity ($v_p$) or, in some cases $N_g$, group velocity ($v_g$) [25, 26]:

$$n = \frac{c}{v} = \frac{kc}{\omega}$$

where $k$ and $\omega$ correspond to the angular wavenumber ( the magnitude of wavevector) and angular frequency, respectively.

The absolute refractive index ($n_a$) is defined when light transitions from a vacuum ($c$) into another medium, whereas the relative refractive index refers to the resulting ($n_r$) when light propagates from one non-vacuum medium ($v_1$) into another ($v_2$):

$$n_r = \frac{v_1}{v_2}$$

By applying Maxwell–Heaviside's equations, a fundamental and simple formula for the relative refractive index can be derived:

$$n = \sqrt{\varepsilon_r \mu_r}$$

where $\mu_r$ denotes the relative magnetic permeability, and $\varepsilon_r$ represents the relative electric permittivity (dielectric constant). For nonmagnetic materials such as nanocellulose, where $\mu_r = 1$, the refractive index simplifies to $n = \sqrt{\varepsilon_r}$. This elegant formula masterfully connects a medium's optical properties to its dielectric properties across any frequency. Moreover, as the formula loudly speaks, the refractive index is influenced by wavelength, leading to a phenomenon known as dispersion.

However, when light propagates through a medium, a portion of it is often attenuated due to mechanisms such as scattering, free-carrier absorption, and phonon generation. This necessitates the introduction of a new term, the complex refractive index ($N_{(\lambda)}$):

$$N_{(\lambda)} = n_{(\lambda)} - ik_{(\lambda)} = \sqrt{\varepsilon_r} = \sqrt{\varepsilon_r' - i\varepsilon_r''}$$

Here, $\lambda$ represents the wavelength, $n_{(\lambda)}$ is the real refractive index, which indicates the phase velocity, and the imaginary part, $k_{(\lambda)}$, represents the attenuation of electromagnetic waves within the material (the extinction coefficient). [27] As the formula suggests, in this scenario, the refractive index becomes a complex function of the light's frequency, where $\varepsilon_r'$ and $\varepsilon_r''$ represents the real and imaginary components of $\varepsilon_r$, respectively. Hopefully, the terms $n$ and $k$ be obtained by analyzing the material's surface reflectance while varying the polarization and angle of incidence. To do so, the relation between these constants, the reflection coefficient ($r$), and the reflection ($R$) is expressed as follows:

$$R = |r|^2 = \frac{k^2 + (1-n)^2}{K^2 + (1+n)^2}$$

In a scenario where $k$ is large, strong absorption occurs and from the above formula results in a reflectance close to unity. Consequently, most of the light is reflected, and attenuation within the medium is significantly high. However, since nanocellulose is not a conductive material [28] and can be considered an insulator—or, in the case of some of its derivatives, a semiconductor—its dispersion can be described using a single-oscillator model. In this model, an external electric field induces forced dipole oscillations within the medium, meaning that the electron shells are displaced and oscillate around the nucleus at a single resonant frequency. [29] Optical properties are often analyzed by examining the frequency dependence of $n$ and $K$ or the real and imaginary components of the relative permittivity, $\varepsilon_r'$ and $\varepsilon_r''$, which can be determined through reflectance, transmittance, ellipsometry measurements, and the following formula:

$$\varepsilon_r' = \frac{N_{a_t}}{\varepsilon_0} a_e' + 1 \qquad and \qquad \varepsilon_r'' = \frac{N_{at}}{\varepsilon_0} a_e'' + 1$$

Where $\varepsilon_0$, $N_{at'}$, $a_e'$, and $a_e''$ represent the vacuum permittivity, the number of atoms per unit volume (which is directly related to the material's density through the molar mass and Avogadro's number), and the real and imaginary components of the electronic polarizability, respectively. For further details, the work of Mistrik et al. is recommended. [30] By using the above formulas, the study of the optical properties of a medium can be conducted by varying the frequency dependence of n and k. Figure 4a illustrates that in a simple single-oscillator model, the real and imaginary components of the complex index reach their maximum values when the normalized frequency is approximately unity, $\frac{\omega}{\omega_0} \approx 1$. However, Figure 4b illustrates that the reflectance and the imaginary component of the complex refractive index reach their maximum values almost simultaneously. This indicates that when reflectance is at its highest, absorption is also strong. Additionally, the plot reveals two distinct regions: when the frequency $\omega$ above $\omega_0$, it corresponds to the anomalous dispersion region, whereas when $\omega$ is below $\omega_0$, it falls within the normal dispersion region. As the graph suggests, below the single resonant frequency $\omega_0$, the refractive index $n$ drops as $\omega$ decreases. Conversely, within the anomalous dispersion region, $n$ decreases as $\omega$ increases. However, this is merely a simplified representation. To achieve a more precise and realistic understanding, we must adopt a quantum mechanical perspective and describe this phenomenon using a quantum dipole oscillator model, in which the photon wave packet excites the oscillator to a higher quantum state. [31]

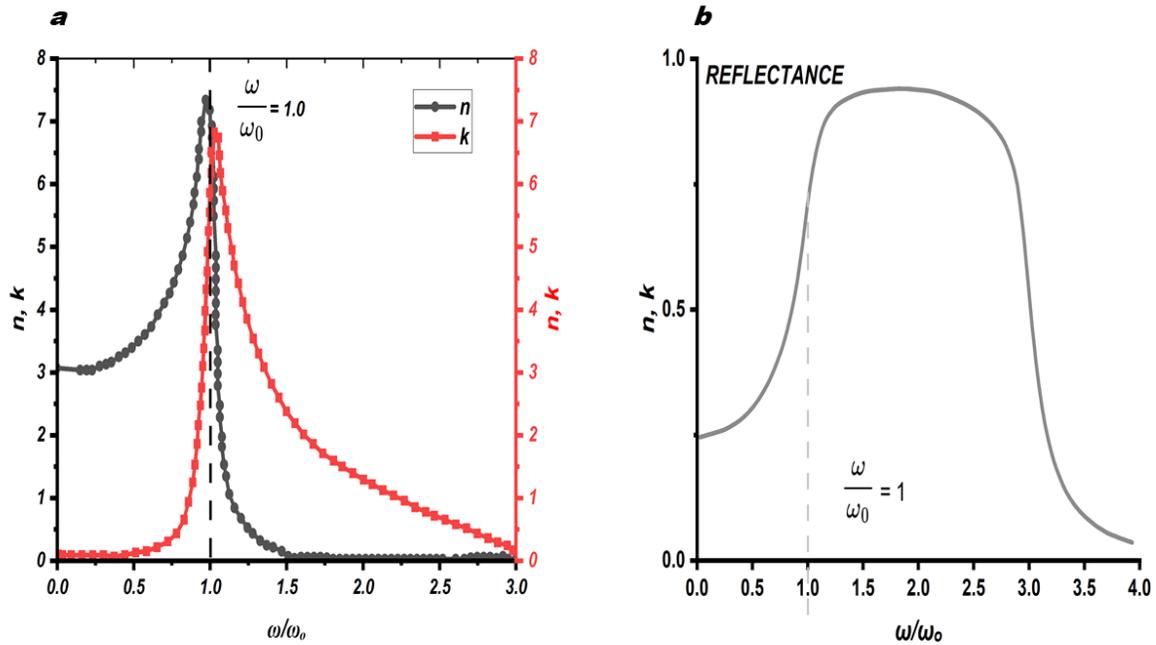

Figure 4. The dipole oscillator model. (a) Refractive index ($n$) and extinction coefficient ($k$) as a function of normalized frequency. (b) Reflectance as a function of normalized frequency. Data from reference[30]

A valuable mathematical tool for deriving the real and imaginary components of relative permittivity is the Kramers–Kronig (K-K) relations. These relations consist of a set of fundamental formulas that establish the connection between the real and imaginary parts of a complex function representing a physical response, utilizing Hilbert transforms. If the response is linear and either the real or imaginary part of the relative permittivity is known, the K-K relations allow us to determine the frequency dependence of the other component across the entire spectrum. The K-K relations are expressed as follows:

$$\varepsilon'_{r(\omega)} = \frac{2}{\pi} P \int_0^\infty \frac{\omega' \varepsilon''_r(\omega')}{\omega'^2 - \omega^2} d\omega' + 1 \quad and \quad \varepsilon''_r(\omega) = \frac{-2\omega}{\Pi} P \int_0^\infty \frac{\varepsilon'_r(\omega')}{\omega'^2 - \omega^2} d\omega'$$

Where P denotes the Cauchy Principal Value of the integral, meaning the integral is evaluated while avoiding the singularity at, π appears as a normalization factor. For further details, reference [32] is recommended.

Several simplified models describe the spectral dependence of the refractive index. Here, we will provide a brief explanation of models and key features that can change optical properties.

The Cauchy dispersion equation is an empirical formula used to describe the variation of the refractive index ($n$) of a transparent material with the wavelength ($\lambda$) of light. It is commonly used in optics for isotropic materials like glass, where normal dispersion is relatively weak in the visible spectrum. The formula is given by:

$$n = a_0 + a_2 \lambda^{-2} + a_4 \lambda^{-4} + a_6 \lambda^6 + \cdots \cdot \lambda > \lambda_h$$

Depending on the chosen method and required precision, the series is truncated at a specific term and the summation $\sum_{j=0}^{\infty} a_{2j}$ represents a material-dependent constant. [33] As evident, this equation is unsuitable for certain nanocellulose-based materials due to their anisotropic nature. Moreover, in addition to rod-shaped cellulose nanocrystals (CNCs) with anisotropic properties that enable their self-organization into chiral nematic liquid crystals (LCs), recent studies have further revealed that spherical cellulose nanocrystals (cellulose nanospheres, CNSs) can also form a chiral liquid-crystalline phase. [34, 35] However, while the Cauchy equation is typically used for isotropic liquids and gases, anisotropic materials require extended versions of the Cauchy equation. For instance, Li et al. [33] proposed a temperature and wavelength dependence formula for both ordinary ($n_o$) and extraordinary ($n_e$) refractive indices using the order parameter:

$$n_e = A_e + \frac{B_e}{\lambda^2} + \frac{C_e}{\lambda^4},$$

$$A_e = A_i + A'_e S,$$

$$B_e = B_i + B'_e S,$$

$$C_e = C_i + C'_e S,$$

Where $A_i$, $B_i$, and $c_i$ represent Cauchy coefficients of the liquid crystal in its isotropic state, while $A'_e = G\lambda_+^2$, $B'_e = G\lambda_+^4$, and $C_e = G\lambda_+^6$.

$$n_o = A_o + \frac{B_o}{\lambda^2} + \frac{C_o}{\lambda^4},$$

$$A_o = A_i - A'_o S,$$

$$B_o = B_i - B'_o S,$$

$$C_o = C_i - C'_o S,$$

Where $A'_o = \frac{G\lambda_+^2}{2}$, $B'_o = \frac{G\lambda_+^4}{2}$, and $C'_o = \frac{G\lambda_+^6}{2}$. However, the order parameter (S), which can also be described using Landau theory of phase transitions in group space [36], represents the degree of orientation or alignment in a system. It indicates how much the system deviates from a higher-symmetry phase. Typically, in phase transitions, the order parameter is zero in the high-symmetry phase (e.g., a liquid) and becomes nonzero in the low-symmetry phase, where the system exhibits greater structural order (e.g., a crystal). The order parameter can be determined through Q-tensors [37] or experimental measurements. For example, when the temperature significantly deviates from the clearing temperature (liquid crystal) and neglecting the internal field, the following approximation can be made [38]:

$$S_{Haller} = \left(1 - \frac{T}{T_c}\right)^\beta$$

Where $\beta$ represents a material parameter. The extended Cauchy equations can be applied to both mixtures and single compounds. As indicated, at a given temperature, both ordinary and extraordinary refractive indices decrease as frequency decreases (normal dispersion). However, when the wavelength extends beyond unity at the microscale, the influence of the second (B) and third (C) terms diminishes. Under these conditions, the ordinary and extraordinary refractive indices converge toward the first term (A), implying frequency independence. Figure 5. shows a comparison between the experimental data and the extended Cauchy equation's prediction for the wavelength-dependent refractive index of 4-cyano-4-n-pentylbiphenyl. [33]

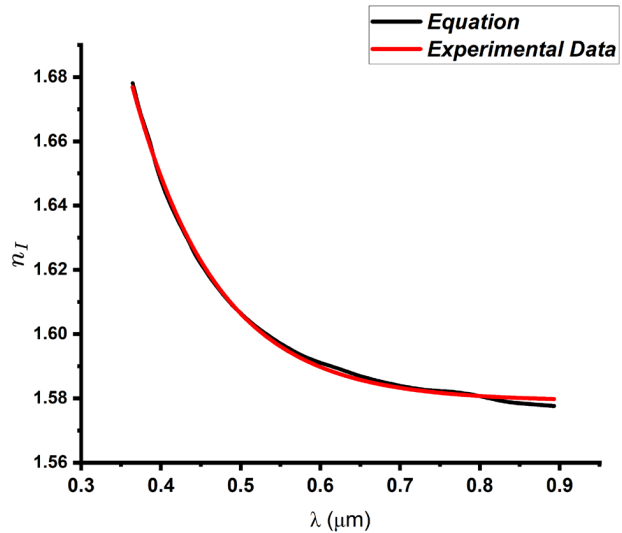

Figure 5. The red line represents the experimental data, while the black line corresponds to the fitting curve obtained using the extended Cauchy equation with the Cauchy coefficients. $n_I$ represents the refractive index in the isotropic state (data sourced from reference[33]).

Li et al [33], also proposed a temperature-dependent formula for a given wavelength, which has shown satisfactory results, expressed as:

$$n_{e(T)} \approx n_i + G'\left(1 - \frac{T}{Tc}\right)^\beta$$

$$n_{o(T)} \approx n_i - \frac{G'}{2}\left(1 - \frac{T}{Tc}\right)^\beta$$

Where $n_i$, $\beta$, and $G' = G\left[\frac{\lambda^2 \lambda_+^2}{\lambda^2 - \lambda_+^2}\right]$ represent the isotropic state refractive index at a given wavelength and the fitting parameters, respectively. The study demonstrated that, at a given temperature, both ordinary and extraordinary refractive indices decrease as the frequency decreases. However, at a fixed wavelength, an increase in temperature results in a slow, gradual increase in the ordinary refractive index, while the extraordinary refractive index shows a decrease. It is important to highlight that molecular polarizability and molecular packing density play a crucial role in the variation of ordinary and extraordinary refractive indices. One of the key relations describing this dependence is the Vuks equation, a semi-empirical model that links microscopic molecular polarizabilities to macroscopic refractive indices for anisotropic media by considering an isotropic internal field in a liquid crystal. This equation is given by [39]:

$$\frac{n_{e,o}^2 - 1}{\langle n^2 \rangle - 2} = \frac{4\pi}{3} N \alpha_{e,o},$$

Where $\alpha_e$ and $\alpha_o$ are extraordinary and ordinary molecular polarizabilities, respectively, and $\langle n^2 \rangle$ represents the mean value of the square of the refractive index, defined as:

$$\langle n^2 \rangle = \frac{(2n_o^2 + n_e^2)}{3},$$

In another study, the temperature dependency of $\langle n \rangle$ can be assessed by analyzing the changes in the temperature-dependent effective molecular density, derived from the experimental data of $n_e$ and $n_o$ :[38, 40]

$$n_o = A + BT - \frac{1}{3}(\Delta n)_o \left(1 - \left(\frac{T}{T_C}\right)\right)^\beta,$$

$$n_e = A + BT + \frac{2}{3}(\Delta n)_o \left(1 - \left(\frac{T}{T_C}\right)\right)^\beta,$$

Here, A and B are determined by fitting a curve that represents the temperature dependence of $\sqrt{\langle n \rangle}$, while $(\Delta n)_o$ denotes the birefringence of the liquid crystal in its crystalline state. However, The values of $(\Delta n)_o$ and β can be obtained through a linear curve fitting approach. Using this equation, the temperature dependence of the refractive indices is illustrated in Figure 6, where symbols represent experimental data, and solid lines indicate the fitting results. [39]

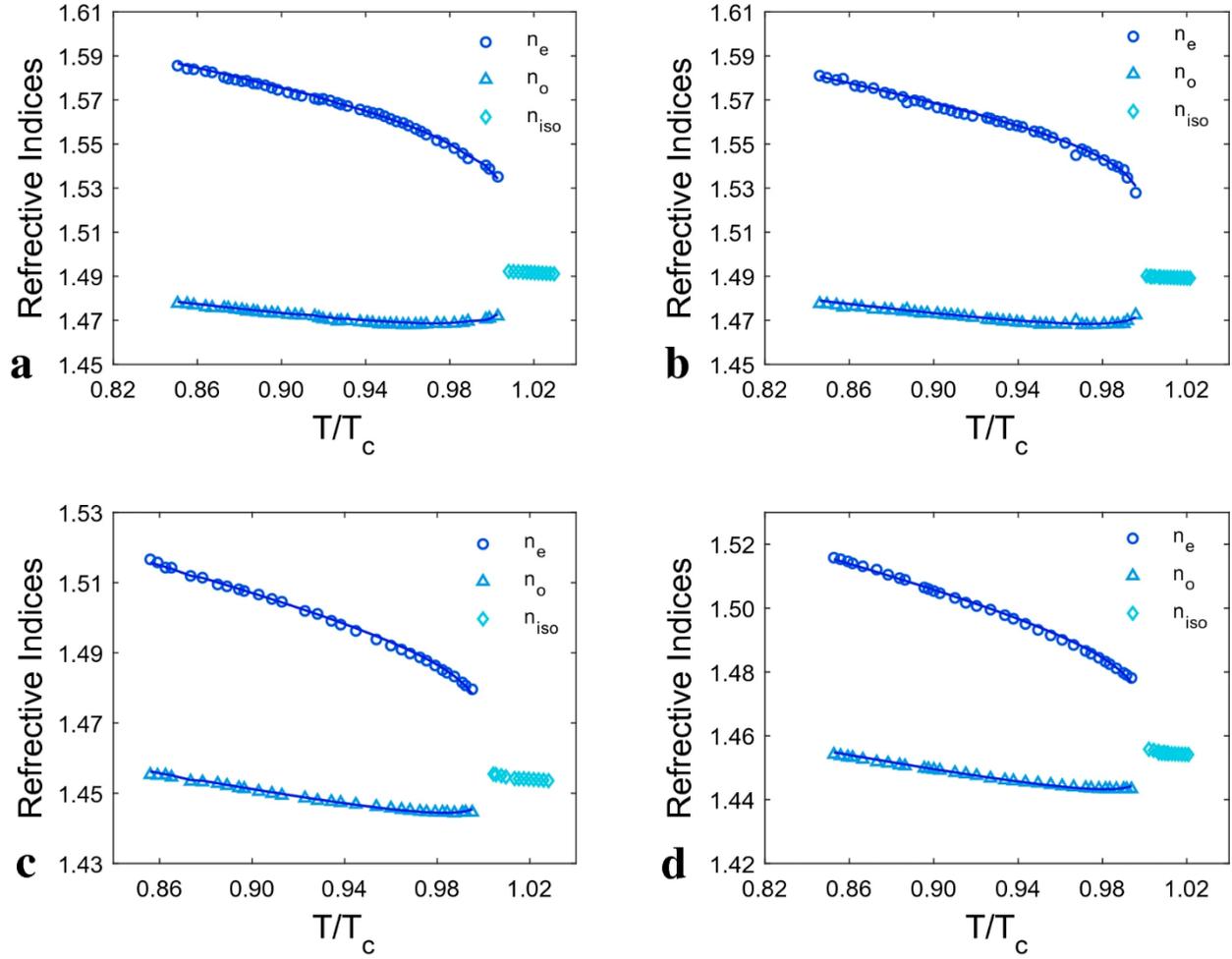

Figure 6. illustrates the temperature dependence of the ordinary, extraordinary, and isotropic refractive indices for (**a**) MAT-131957 (LC-I), (**b**) MAT-131958 (LC-II), (**c**) ZKC5102-LA (LC-III), (**d**) ML-0682 (LC-IV), where symbols represent experimental data, and solid lines denote the fitting results. Copyright 2022 the authors. Published by Nature under a Creative Commons CC BY License. [39]

Table 1. presents a comparison between the experimental birefringence values and the corresponding fitting results [39])

| Liquid Crystal | $\Delta n_{Ex}$ | $\Delta n_{Fit}$ |
|---|---|---|
|  |  |  |
| MAT-131957 (LC-I) | 0.1070 | 0.1403 ± 0.47% |
| MAT-131958 (LC-II) | 0.1066 | 0.1364 ± 1.01% |
| ZKC5102-LA (LC-III) | 0.0760 | 0.0854 ± 1.18% |
| ML-0682 (LC-IV) | 0.630 | 0.0836 ± 0.37% |

In a separate study, Yang et al.[41] investigated the correlation between temperature and delay in silica optical fibers by considering the fundamental relationship between the optical fiber refractive index and delay, given by $n_{(T)} = \frac{\tau c}{L}$, where c is the speed of light in a vacuum, L is the length of optical fiber, $\tau$ represents the delay. Furthermore, to incorporate the influence of an external electric field (E), the relationships among electric susceptibility ($\chi$), temperature (T), density ($\rho$), and molecular polarizability ($\alpha$) in a non-magnetic medium can be expressed as follows:

$$n = \sqrt{\varepsilon_r} = \sqrt{1+\chi} = \sqrt{1 + \frac{N\alpha_0}{\varepsilon_0} + \frac{\rho\eta C(T-T_o)}{\varepsilon_0 E^2}}$$

Where $\chi, N, C, \eta, \rho, and\ E$ represent the electric susceptibility, the density of electric dipoles, the specific heat coefficient, the proportionality coefficient, the medium density, and the external magnetic field, respectively.

This section provides a brief overview of one of the models used to calculate the refractive index of materials with different properties, structures, and functional groups, based on parameters such as band gap and electronegativity. In a study by Herve et al., a refractive index model based on oscillatory theory has been proposed:

$$n = \sqrt{1 + \left(\frac{A}{E_g + B}\right)^2}$$

Where $E_g$ represents the energy gap, and $A = 13.6\ ev$ and $B = 3.4\ ev$ are the constants. However, this equation is not applicable to materials with a high-energy band gap ($E_g \geq 4$). [42] Moreover, as demonstrated in previous research, a correlation exists between the refractive index and optical electronegativity for oxides, semiconductors, and insulators.[43-45] This suggests the possibility of developing an electronegativity-responsive model for nanocellulose-based materials to predict refractive index variations in binary systems.

Furthermore, surface roughness plays a crucial role in determining optical properties. Here, we provide a description of surface roughness and briefly introduce one of the effective models used to characterize it. To analyze surface topography, techniques such as the Fourier transform can be employed, converting surface height data from the spatial domain (position) to the frequency domain. By applying the Fourier transform to a function that quantifies the correlation between surface height z(x,y) at different points—known as the Surface Autocovariance Function (ACF)—we obtain a function that describes the distribution of surface roughness across various spatial frequencies, referred to as the Power Spectral Density (PSD), which a 2D-isotropic PSD with a polar symmetry is defined by averaging the 2D-PSD over all azimuthal directions:

$$PSD(f) = \frac{1}{2\pi} \int_0^{2\pi} PSD(f,\psi)\, d\psi,$$

Where a simple solution is $f^2 = f_x^2 + f_y^2$ and $\tan(\psi) = \frac{f_y}{f_x}$. here, $f_x$ and $f_y$ are the lateral surface spatial frequencies.

The investigated surface area and the instrumental resolution constrain each roughness measurement to a specific spatial frequency range. For more information about PSD and its applications, this reference is [46] recommended. Here, we use the Root Mean Square (RMS) method to quantify isotropic surface roughness. RMS roughness ($\sigma$) is commonly defined as the square root of the standard deviation of z(x,y) from its mean value and is given by:

$$\sigma = \left[2\pi \int_{f_i}^{f} PSD(f) f \, df\right]^{\frac{1}{2}},$$

The integration is determined based on the measurement technique used. Now, we can define scattering for an isotropic surface with polar roughness, referred to as Angle-Resolved Scattering (ARS), which is expressed as [47]:

$$ARS(\theta_s) = \frac{\Delta P_s(\theta_s)}{P_i} \frac{1}{\Delta \Omega_s},$$

Where $P_i$ represents the incident power, $\Delta P_s(\theta_s)$ denotes the scattered power at the angle $\theta_s$, and $\Delta \Omega_s$ is the scattering solid angle normalized to the incident angle. Angle-resolved scattering is widely employed in the study of nano-based and nanocellulose-based materials [48] using techniques such as Angle-Resolved Raman Scattering [49] and is often combined with other methods like Small-Angle X-ray Scattering (SAXS).[50]

One of the effective scattering models applicable when the ratio of RMS roughness ($\sigma$) to incident wavelength ($\lambda$) is less than either one-hundredth ($\frac{\sigma}{\lambda} < 0.01$)[51] or five-hundredth ($\frac{\sigma}{\lambda} < 0.05$)[52] is the Rayleigh-Rice (RR) theory. The Rayleigh-Rice vector theory offers a rigorous analytical solution to Maxwell's equations for the scattering of electromagnetic waves from smooth surfaces. Based on perturbation theory, it provides an exact solution within its domain of validity, expressed as:

$$ARS(\theta_s) = \left(\frac{4\pi}{\lambda^2}\right)^2 \gamma_i \gamma_s^2 \theta_s \, Q \, PSD(f),$$

Here, $\gamma_i = \cos \theta_i$ and $\gamma_s = \cos \theta_s$ are the optical factors, defined as the cosines of the incident and scattering angles, respectively. Additionally, $Q$ is an optical factor that encompasses information related to polarization, scattering, angle of incidence, and the dielectric function, commonly referred to as the angle-dependent polarization reflectance.[47] Furthermore, the relationship between the scattering angle and spatial frequency is given by $f = \frac{|\sin\theta_s - \sin\theta_i|}{\lambda}$. In the smooth regime, which merely consist of the first diffraction orders of the surface spectral components. The plots in Figure 7 compare different scattering models at various surface roughness levels to evaluate their effectiveness. For further details, refer to the work of Schröder et al.[47]

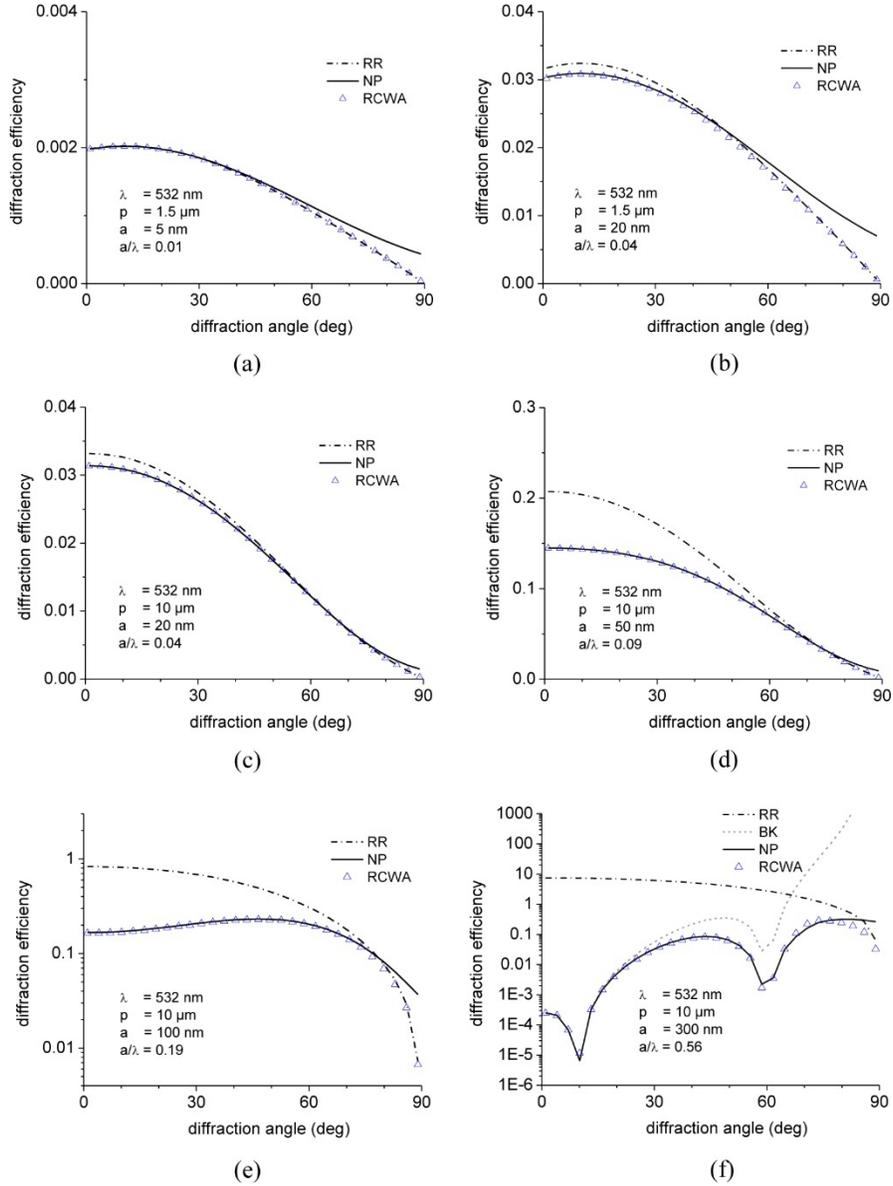

Figure 7. The scattering efficiency of first-order diffraction for sinusoidal phase gratings, each distinguished by unique periods $p$ and amplitudes $a$, was meticulously analyzed at a wavelength of 532 nm. By precisely manipulating the angle of incidence, diffraction angles spanning the entire range from 0° to 90° were successfully attained. Here, RR, BK, RCWA, and NP stand for the following scattering models: Rayleigh-Rice (RR) theory[53], Beckmann-Kirchhoff (BK) theory[54], the Rigorous Coupled Wave Analysis[55], and the non-paraxial diffraction theory (in this special

case, the Harvey-Shack theory)[56], respectively. Copyright 2011 the authors. Published by Wiley under a Creative Commons CC BY License. [47]

In another study, Reimer et al[57]. utilized the transfer matrix method to calculate the propagation of the refractive index $n(\lambda)$ for cellulose fibers by fitting the resulting transmittance curves to a model described by the Sellmeier equation:

$$n(\lambda) = \sqrt{\frac{A + B\lambda}{\lambda - C}}$$

Here, the Sellmeier coefficients $A, B, and\ C$ were independently refined. However, by accounting for scattering losses due to reflection and absorption $T_{\alpha,r}$ and applying the transfer matrix method, the following formula was used to approximate the scattering losses:

$$T_{tot} = T_{a,r} - \frac{C_s}{\lambda^\alpha}$$

Where $\alpha$ represents the refined size-dependent dispersion component, while $C_s$ denotes the refined apparent concentration of scatterers. After determining the refractive index curve, the Abbe number $v_D$ and the Abbe diagram can be calculated using the equation below. The Abbe number represents the constringence of a transparent material, meaning it provides a useful approximation of the material's dispersion. Lower Abbe number values indicate higher dispersion and refractive index. However, attenuation also plays a crucial role and is primarily influenced by factors such as wavelength, temperature, and filament length.[57, 58]

$$v_D = \frac{n_D - 1}{n_f - n_c}$$

Where $n_D$, $n_f$, and $n_C$ represent the refractive indices at D-, F-, and C-lines, respectively. Reimer et al. utilized Fraunhofer lines for these values; however, they can also be determined using alternative approaches, such as the Sellmeier equation.[57]

1.1) AI-Driven Models for Predicting Optical Properties

Undeniably, one of the critical challenges in material science is the integration of artificial intelligence in material discovery, screening, and property prediction. However, the limited number of scientific studies utilizing these approaches to advance the discovery of optical properties in nanocellulose-based materials is evident. Conversely, as demonstrated in the research, a machine learning approach can achieve speeds up to 16 times faster than the conventional finite-difference time-domain (FDTD) method for obtaining optical properties.[59] This remarkable acceleration allows for the evaluation and optimization of a vast number of nanomaterial candidates. For instance, one study utilized a trained machine learning model, derived from computationally generated spectra of chiral plasmonic structures, to predict spectra for an extensive dataset comprising 28,000 structures with unprecedented efficiency.[60] Here, we explore some of the models that have been applied to polymeric and nanomaterials for predicting optical properties. In a study, Signori-Iamin et al.[61] employed a model incorporating Random Forests (RF), Linear Regression (LR), and Artificial Neural Networks (ANN) to predict the aspect ratio of micro/nanocellulose fibers. The trained Artificial Neural Networks achieved an $R^2$ value of 0.96 with a Mean Absolute Percentage Error (MAPE) of 4.54%. Additionally, Random Forests and Linear Regression yielded correlation coefficients of 0.93 and 0.95, respectively. Although this research did not directly investigate the optical properties of nanocellulose, the findings are still relevant. As discussed earlier, factors such as density, aspect ratio, and the orientation of nanocellulose fibers significantly influence the optical properties of these materials. Therefore, the data obtained in this study can also be utilized for this purpose. In another study, researchers Mohsin Khan et al.[62] developed four quantitative structure-property relationship (QSPR) models with $R^2$ values of 0.895, 0.895, 0.897, and 0.896, and mean absolute errors of 0.004, 0.0041, 0.0052, and 0.0051 for different models, respectively. These models were constructed using 2D descriptors based on polymers' monomer units to predict the refractive indices of a dataset comprising 221 different organic polymers. The dataset was divided into a training set (154 compounds) and a test set (67 compounds) using the Kennard–Stone method.[63] A double cross-validation approach[64] was employed to identify six key descriptors, followed by the application of partial least squares regression[65] to build the predictive model. To avoid the complexities of polymeric interactions, instead of using the full polymer structures, the study derived descriptors solely from single monomer units with end-capped hydrogen atoms.[66] According to the authors, the selected models demonstrated satisfactory performance based on both internal and external validation metrics, including mean absolute error-based criteria[67] and Golbraikh and Tropsha's criteria[68]. Figure 8 presents the overall workflow diagram.

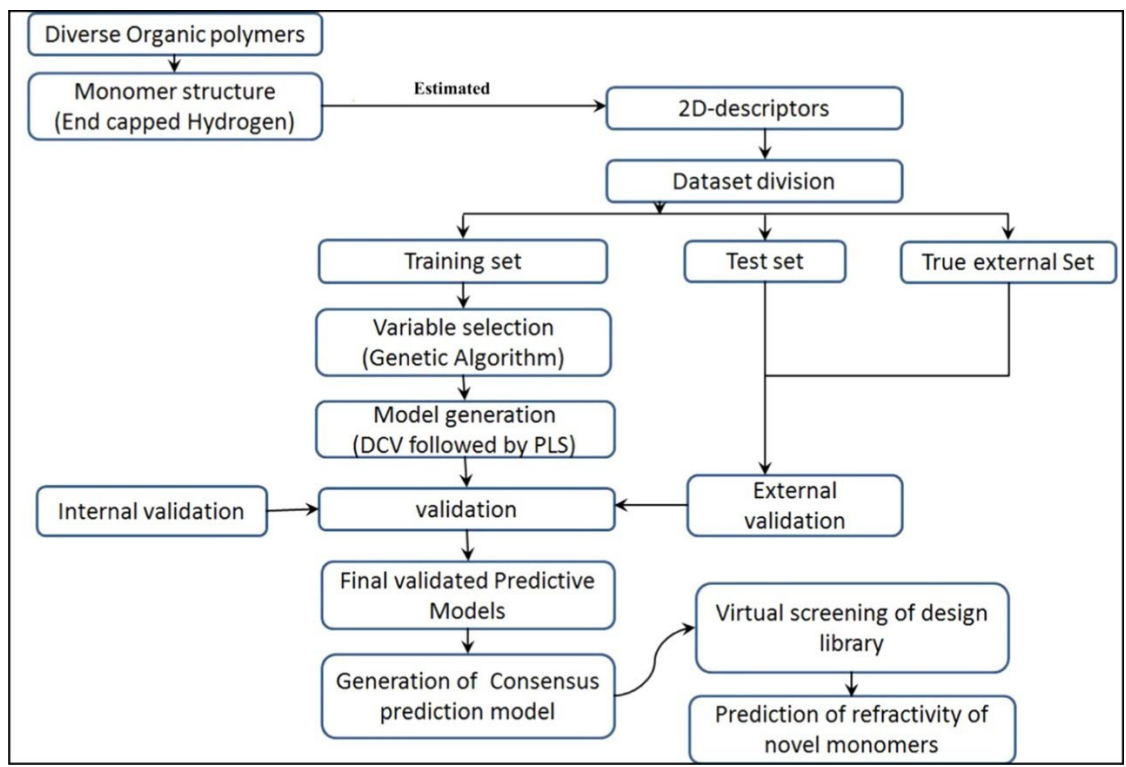

Figure 8. Workflow diagram of the methodology used. Copyright 2018 by the authors. Published by ACS under a Creative Commons CC BY License.[62]

In a separate study, Stein et al.[69] developed three autoencoding algorithms to analyze the measured optical properties of metal oxides, enabling the extraction of essential data from images and spectra to predict the fundamental structural characteristics underlying experimental materials science data. The dataset consisted of 178,994 distinct material samples, spanning 74 different composition spaces, including over 80,000 unique quinary oxide and 67,000 quaternary oxide compositions. The data was used to train and validate three different models for mapping relationships between sample images and absorption spectra. The optical band gap energies autogenerated from patterns were as accurate as traditional methods based on full transmission and reflection spectra measurements, with a mean absolute error of 180 meVOne of the models developed is the Absorption Spectra Prediction Model (ASPM). This model encodes extracted image data—such as morphology and color variations—into a compact latent space representation, reducing the dimensionality from 12,288 dimensions (corresponding to a 64 × 64 red-green-blue image) to 100 dimensions. The encoded information is then decoded to predict absorption spectra using a hybrid dense and convolutional deep neural network for training. As shown in Figure 9a, the high Pearson correlation coefficients and $R^2$ values, along with the low relative mean absolute error and root mean square deviation, indicate that this model can effectively predict absorption spectra and optical band gaps. Additionally, Figure 9b demonstrates the accuracy of the model by comparing the ground truth absorption spectra (green) with the predicted spectra (black). The predictions are precise enough to capture fine spectral features, such as local maxima caused by sub-bandgap absorption or thin-film interference. As noted by the authors, these fine spectral details are often beyond the sensitivity of the original RGB sensors and remain hidden even to expert analysis. This highlights the superhuman analytical capabilities of the machine learning models, which can extract intricate features that are imperceptible to the human eye. [69]

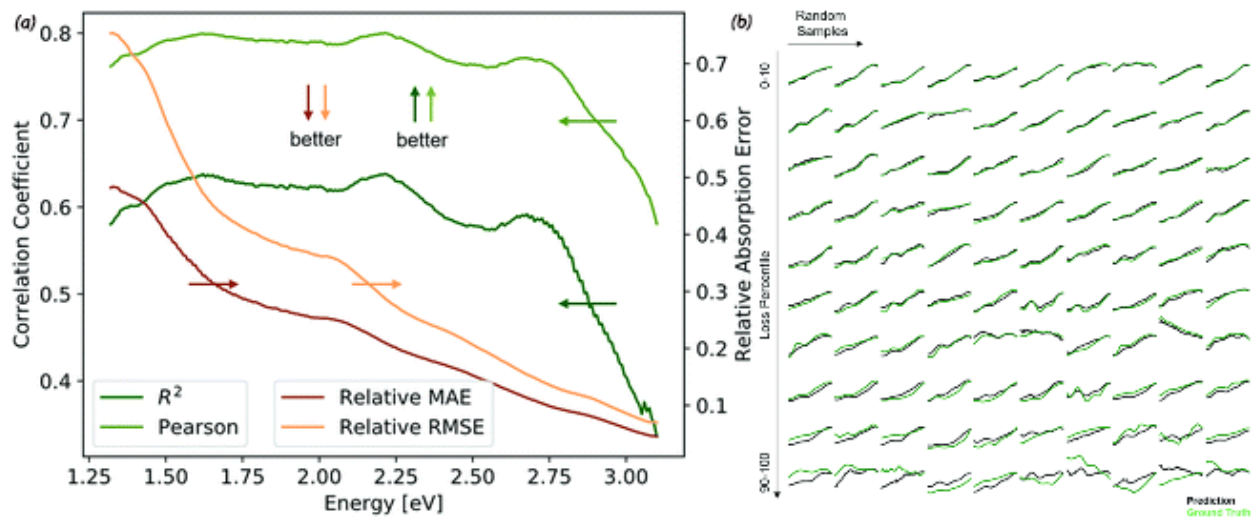

Figure 9. (a) A plot illustrating the Pearson correlation coefficient $R^2$, relative mean absolute error, and relative root mean square error as functions of the predicted energy for spectra. The Pearson correlation coefficient and $R^2$ maintain above values of 0.75 and 0.6, respectively, up to 2.75 eV. Beyond this energy range, the consistently high optical absorption of ultraviolet waves likely

contributes to reduced variations in the data, leading to lower relative mean absolute error and relative root mean square error. (b) Ten randomly selected examples are presented from each mean squared error (MSE) loss decile, with low-loss deciles, characterized by finely preserved features, displayed at the top, while high-loss deciles, exhibiting lower predictability, appear at the bottom. Notably, within the 0–80 percentile range of MSE loss, the model effectively predicts both the overall spectral shape as well as intricate details, such as local maxima. Copyright 2019 by the authors. Published by RSC under a Creative Commons CC BY License.[69]

2) Nonlinear Optical Phenomena and Chiroptical Responses in Nanomaterials

In most everyday experiences, light-matter interactions are governed by linear optics. This implies that the interaction between incoherent light and matter can be analyzed independently, with individual interactions being superimposed to obtain a complete result, such as in the computation of a rainbow. [70]

While polarization $P(r,t)$ is linearly proportional to the applied electric field $E(r,t)$ for small field strengths, linear optics models fail when the applied field becomes comparable to the interatomic electric field ($10^5$-$10^8$ V/m). [71] In such cases, additional terms must be introduced, as shown in the equation below:

$$P_{(r,t)} = \varepsilon_0 \left( \chi \cdot E(r,t) + \chi^{(2)} : E(r,t)E(r,t) + \chi^{(3)} : E(r,t)E(r,t)E(r,t) \right)$$

Here, $\chi, \chi^{(2)}, \text{and } \chi^{(3)}$ represent rank-2, rank-3, and rank-4 tensors, respectively. However, ':' indicates the inner product. [72]

However, when dealing with nanomaterials, the assumption of linear, homogeneous, and isotropic materials is no longer valid, making nonlinear effects more prominent. On the other hand, many nanocellulose materials, such as cellulose nanocrystals (CNCs) and bacterial nanocellulose [73], exhibit susceptibility to second harmonic generation due to their non-centrosymmetric macromolecular structure, high degree of assembly, and crystallinity [74], Although enhancing nonlinear optical properties often requires doping or incorporating additional materials, such as metals. In recent decades, second harmonic generation from metals such as gold and silver has garnered significant attention due to delocalized electron oscillations at metal-dielectric interfaces, known as surface plasmon resonance (SPR). [75]SPR enhances nonlinear optical properties by enabling the concentration of electromagnetic waves in regions smaller than the diffraction limit. [76] Furthermore, there are two principal types of SPR, both of which can be utilized to enhance weak nonlinearity at metal-dielectric interfaces: First, Localized Surface Plasmon (LSP)[77] – A stationary electromagnetic mode confined to the surface of subwavelength-sized metal particles embedded in a dielectric medium—Latter, Surface Plasmon Polariton (SPP)[78] – An electromagnetic mode that propagates along a metal-dielectric interface. In other words, nonlinearity in dielectrics can be enhanced through field amplification associated with the excitation of surface plasmon resonance in plasmonic metals. In general, a surface polarization at frequency $\omega_3$ is generated due to a second-order nonlinear interaction of two excitation waves with frequencies $\omega_1$ and $\omega_2$, which its equation is given by [71, 79]:

$$P_s^{(2)}(r, \omega_3) = \varepsilon_0 \delta(r - r_s) \chi_s^{(2)}(\omega_1, \omega_2, \omega_3) : E(r, \omega_1) E(r, \omega_2)$$

Where $\delta$ and $r_s$ represent a Dirac delta function and position vector of the surface, respectively. In a study, El-Nagger et al. designed polymeric nanocomposites to enhance both linear and nonlinear optical properties. They prepared Polyvinyl alcohol (PVA)/carboxymethyl cellulose (CMC)/polyethylene glycol (PEG) films containing varying weight percentages (0, 1, 3, 5, 10 wt%) of ZnS/V nanoparticles using the casting technique. For the pure polymeric nanocomposite, which was metal-free, a tailored direct and indirect optical band gap ranging from 5.63 to 5.28 eV was observed. In contrast, for the polymeric blend doped with 10% ZnS/V, the band gap decreased to 3.64-2.77 eV. Moreover, the ZnS/V nanoparticles significantly affected the fluorescence spectrum intensity, showing a notable change in its optical properties. [80] In another study, in an effort to fabricate a flexible nanocomposite film, Atta et al.[81] prepared flexible methyl cellulose/polyaniline (PANI) films. Different amounts of PANI (1.0 wt% and 3.0 wt%) were incorporated, followed by the addition of 0.5 wt% and 1.0 wt% of $AgNO_3$ over a specific period. Subsequently, 30 mL of a reducing agent, $NaBH_4$ (2 mM), was introduced to synthesize silver nanoparticles, aiming to enhance both linear and nonlinear optical properties. However, the films were prepared using the casting method, and their optical properties were analyzed using a UV-Vis spectrometer within the wavelength range of 200 nm to 1100 nm. Figure 10 and Table 2 present variations in the optical properties of the methylcellulose, polyaniline, and silver nanoparticle nanocomposites. [81]

Table 2. shows the values of Urbach energy, optical band gap, and absorption edge for different blends of MC, PANI, and AgNPs. Data from Atta et al [81]

| Samples' Code | Urbach Energy (eV) | Optical Band Gap (eV) | Absorption Edge (eV) |
| --- | --- | --- | --- |
| Pristine MC | 0.51 | 5.76 | 5.42 |
| MC/1%PANI | 1.38 | 5.06 | 3.70 |
| MC/3%PANI | 1.52 | 5.01 | 3.14 |
| MC/3%PANI/0.5%Ag | 1.92 | 4.41 | 2.38 |
| MC/3%PANI/1%Ag | 2.02 | 4.27 | 1.93 |

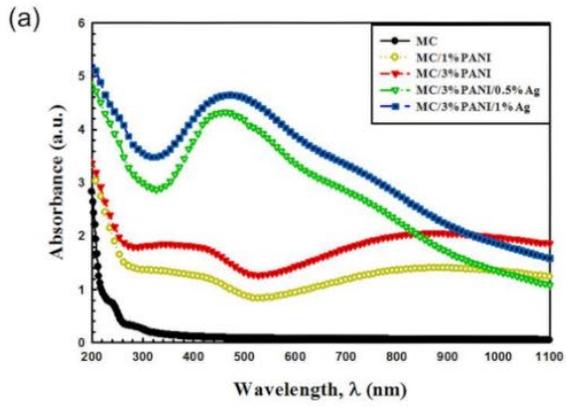
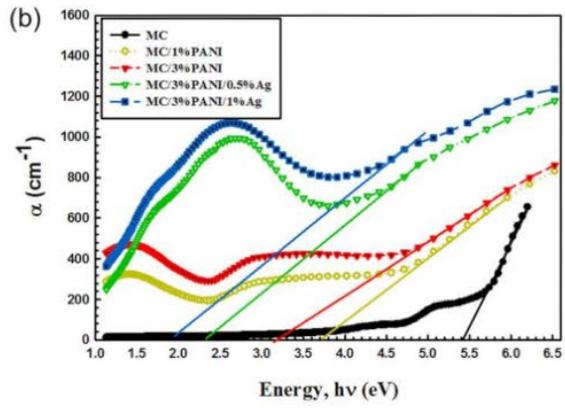
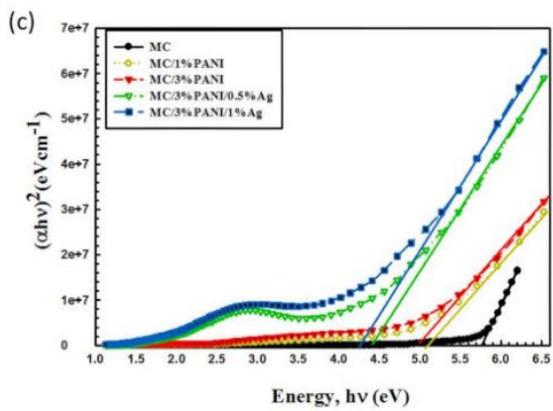
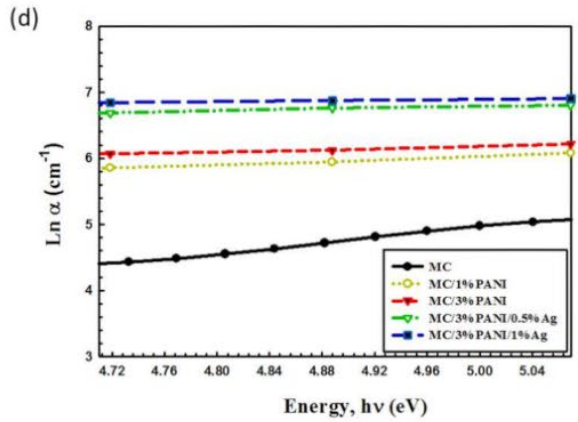
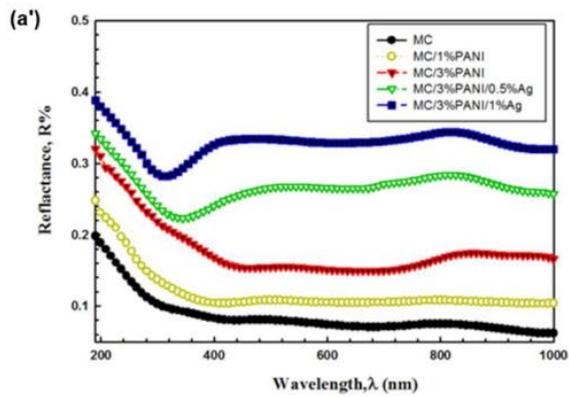
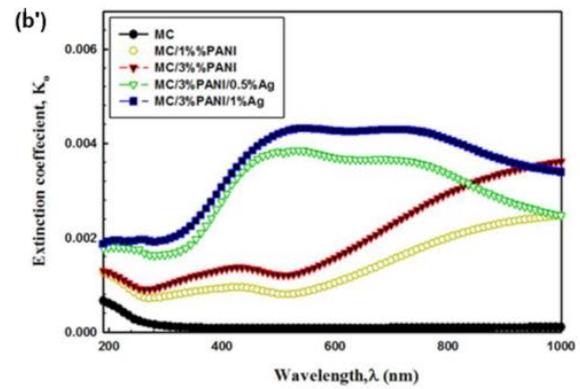
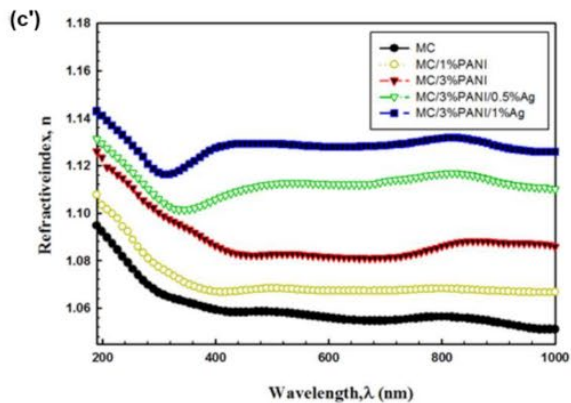
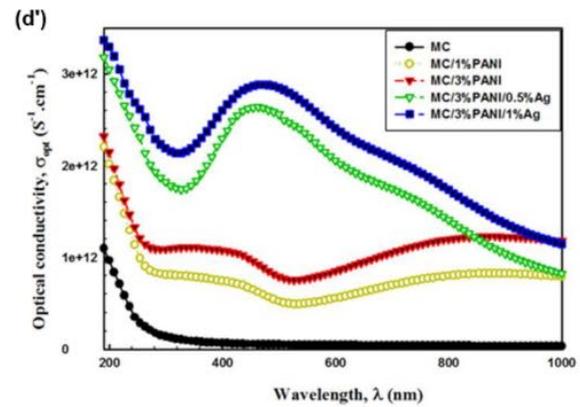

Figure 10. presents various optical properties of different combinations of methylcellulose (MC), polyaniline (PANI), and silver nanoparticles (AgNPs), including: (a) UV-Vis absorption spectra, (b) absorption coefficient α, (c) the relation between $(\alpha h\nu)^2$ against photon energy (hν), (d) absorption coefficient lnα versus photon energy, (a') the reflectance, (b') the extinction coefficient k, (c') the refractive index n, and (d') the optical conductivity σ opt; as a function of wavelength λ. Copyright 2021 the authors. Published by MDPI under a Creative Commons CC BY License. [81]

Figure 10a reveals an additional peak following the incorporation of AgNPs, which is attributed to surface plasmon resonance. Furthermore, the blending of PANI and MC results in the overlapping and enhancement of their respective absorption peaks. Figure 10b illustrates a decrease in the absorption edge, consistent with the data presented in Table 2, when PANI and AgNPs are introduced into the MC polymer. This observation supports the theoretical concept that an increase in excitonic interactions occurs. [82] Moreover, modifying the MC matrix—thereby altering localized states, phononic interactions, and the degree of disorder—leads to the broadening and shifting of the absorption edge. [83] However, Figure 10c supports previous theories, demonstrating a reduction in the optical band gap with the incorporation of PANI and AgNPs. This result confirms that the reduction is attributed to the degree of disorder [84], charge transfer complex formation [85], and defects induced by localized states. On the other hand, Urbach energy gradually increases as the degree of disorder and defects rises. Additionally, the incorporation of PANI and AgNPs facilitates a redistribution of band states, enabling additional tail-to-tail transitions as shown in Figure 10d. The addition of PANI and AgNPs led to an increase in reflectance due to enhanced packing density, the functional groups of PANI, changes in chain orientation [86], modifications in the electronic structure [87], and the intrinsic properties of AgNPs.

Figure 10b' presents the extinction coefficient plotted against wavelength, revealing an increase in the extinction coefficient with the addition of AgNPs and PANI to the MC polymer chain. This increase is attributed to several factors, including the surface plasmon resonance (SPR) of AgNPs, structural modifications in the polymer, and a higher rate of exciton production due to the introduction of new energy levels within the optical band gap. [88] As previously discussed, the incorporation of AgNPs and PANI induced structural modifications in the MC polymer, including changes in polarizability, density[89], the density of localized states [90], and exciton generation [91]. These factors, in turn, contributed to variations in the refractive index and optical conductivity, as shown in Figure 10c' and Figure 10d', respectively[81]. In a separate study, Ramdzan et al.[92] developed a nanocrystalline cellulose (NCC)/poly(3,4-ethylenedioxythiophene) (PEDOT) thin film for plasmonic sensing in mercury ion detection. The optical band gap of the blend was 4.082 eV, which was slightly lower than that of the individual thin films (NCC thin film = 4.101 eV and PEDOT thin film = 4.122 eV), as shown in Figures 11b, 11c, and 11d. Furthermore, Figure 12 illustrates an increase in the incident angle from 54.0099° to 54.4086° as the mercury concentration rises from 0.001 ppm to 0.1 ppm. This shift occurs because mercury ions occupy functional group sites on the NCC/PEDOT thin film, causing the surface plasmon resonance (SPR) signal to shift to higher values.

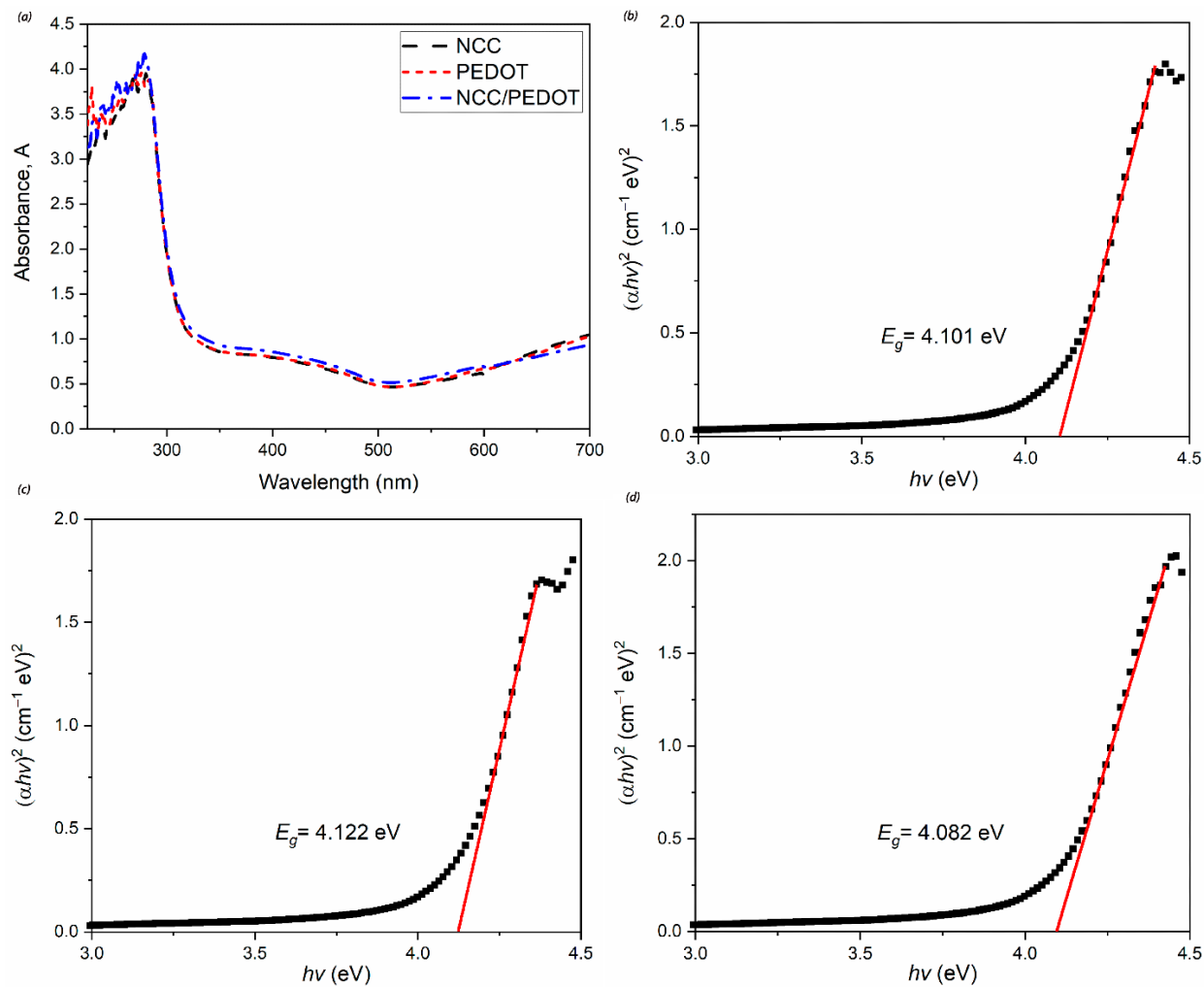

Figure 11. (a) UV-Vis spectra of NCC, PEDOT, and NCC/PEDOT, (b) optical band gap of the NCC thin film, (c) optical band gap of PEDOT, and (d) optical band gap of NCC/PEDOT, determined by extrapolating the straight-line plot on the x-axis using Tauc's equation[93]. Copyright 2021 the authors. Published by MDPI under a Creative Commons CC BY License.[92]

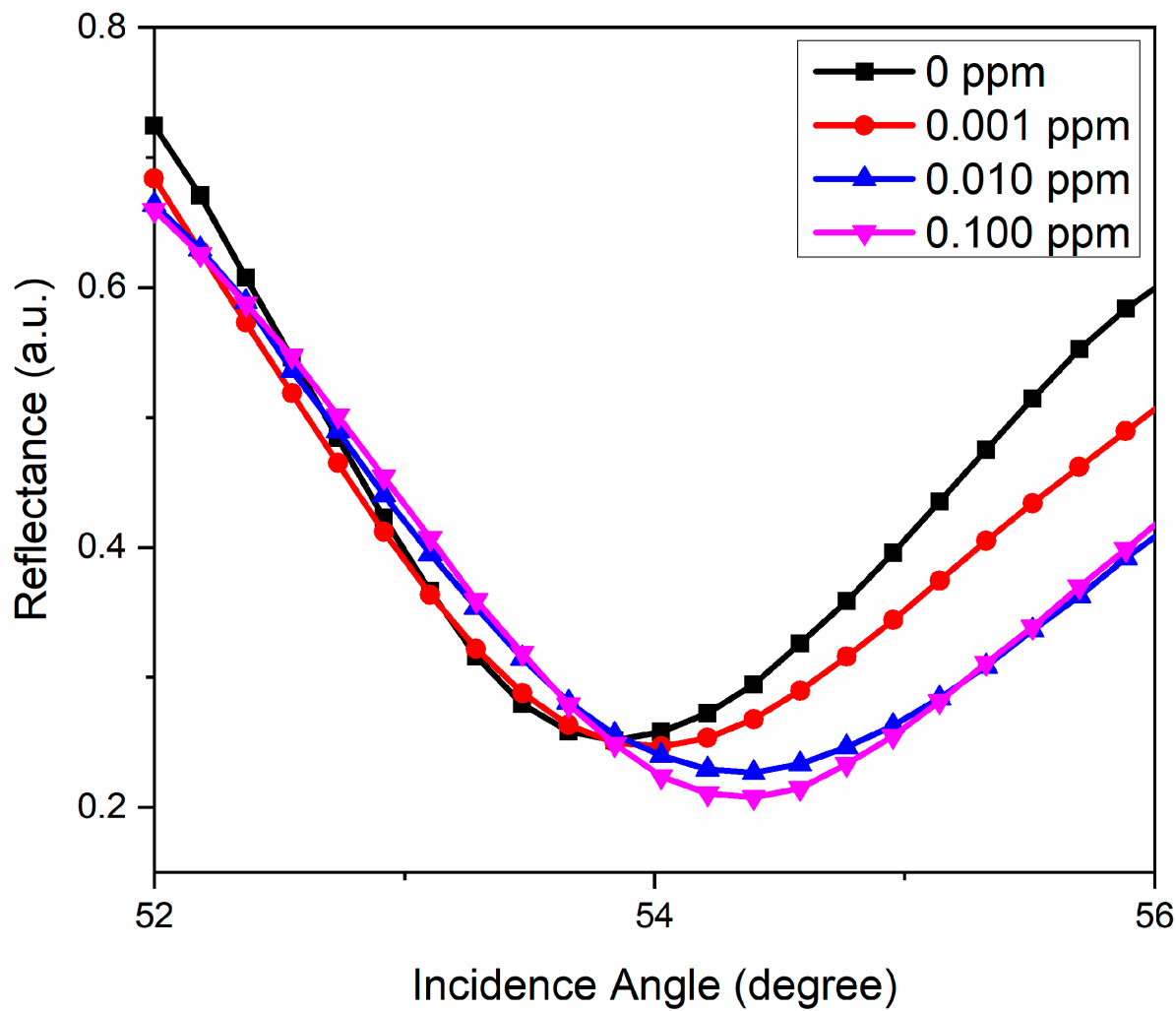

Figure 12. Reflectance of the NCC/PEDOT thin film as a function of the incidence angle for deionized water (black) and different concentrations of mercury ion solution. Copyright 2021 the authors. Published by MDPI under a Creative Commons CC BY License.[92]

One of the potential applications of nanocellulose-based materials is in polarization information encryption and optical anti-counterfeiting, which rely on the emission of left- and right-handed circularly polarized luminescence (CPL). When a material—particularly chiral ones—emits right- and left-circularly polarized light unequally, it gives rise to the phenomenon of circularly polarized luminescence. Chirality primarily arises from multiple levels of the material's structural hierarchy.

However, CPL can also occur in composite systems, where a luminophore is combined with a chiral host.[94] In periodic crystals, this effect has been attributed to periodic variations in dielectric constants. [95] There is a strong correlation between the magnetic component of electromagnetism and CPL, as its occurrence is often associated with allowed magnetic dipole transitions. To determine the strength of CPL, the Kuhn asymmetry factor $(g_{Lum})$[96] can come in handy as its formula follows:

$$g_{Lum} = 2\left(\frac{I_l - I_r}{I_l + I_r}\right) = \frac{4R}{D} \approx 4\frac{|\vec{m}|}{|\vec{\mu}|}\cos\theta$$

Here, $R, D, |\vec{m}|, |\vec{\mu}|, \theta, I_r,$ and $I_l$ represent the following parameters: the rotational strength, dipole strength, magnetic dipole moment strength, electric dipole moment strength, angle between the magnetic and electric dipole moments, emission intensities of right, and left-polarized light, respectively.[97] Two significant structural features that can act as a double-edged sword in optical properties are aromaticity and helical structures, both of which are often associated with π-π stacking. While π-π interactions can enhance conjugation, they can also lead to non-radiative decay, ultimately resulting in decreased quantum yield and fluorescence quenching. [98] However, it is worth noting that nanocellulose-based materials, which do not inherently possess π-systems, can circumvent this particular challenge, making them advantageous for applications requiring high fluorescence efficiency. The unique luminescence pathway of nanocellulose-based materials, particularly cellulose nanocrystals (CNCs), makes them ideal candidates for use as chiral hosts—due to their ability to self-assemble into a helical structure—or as primary fluorophores in circularly polarized luminescence applications. For instance, the chiral nematic structure of CNCs enables them to selectively transmit right-handed circularly polarized light while reflecting left-handed circularly polarized light. [99] Researchers Wang et al. developed multi-stimuli-responsive circularly polarized luminescence (CPL) composite films by incorporating cellulose nanocrystals

(CNC), polyethylene glycol (PEG), and achiral luminescent Eu-containing polyoxometalates (Na$_9$EuW$_{10}$O$_{36}$·32H$_2$O (EuW$_{10}$)) through solvent evaporation at room temperature. To tune the photonic band gap (PBG), they varied the mass fractions of CNC and PEG as follows: 100/0, 90/10, 80/20, 70/30, and 60/40. Their findings indicated that PEG, due to its oxygen-bearing functional groups, along with CNC's superb hydrophilicity and hydrogen bonding capacity, influenced the helical pitch in a manner similar to water. The pitch increased linearly from 275 nm to 445 nm, demonstrating that at low PEG concentrations, the CNC-PEG composite retained its helical structure and birefringence, as confirmed by polarized optical microscopy (POM), scanning electron microscopy (SEM), other research (Figure 13). [100-102]

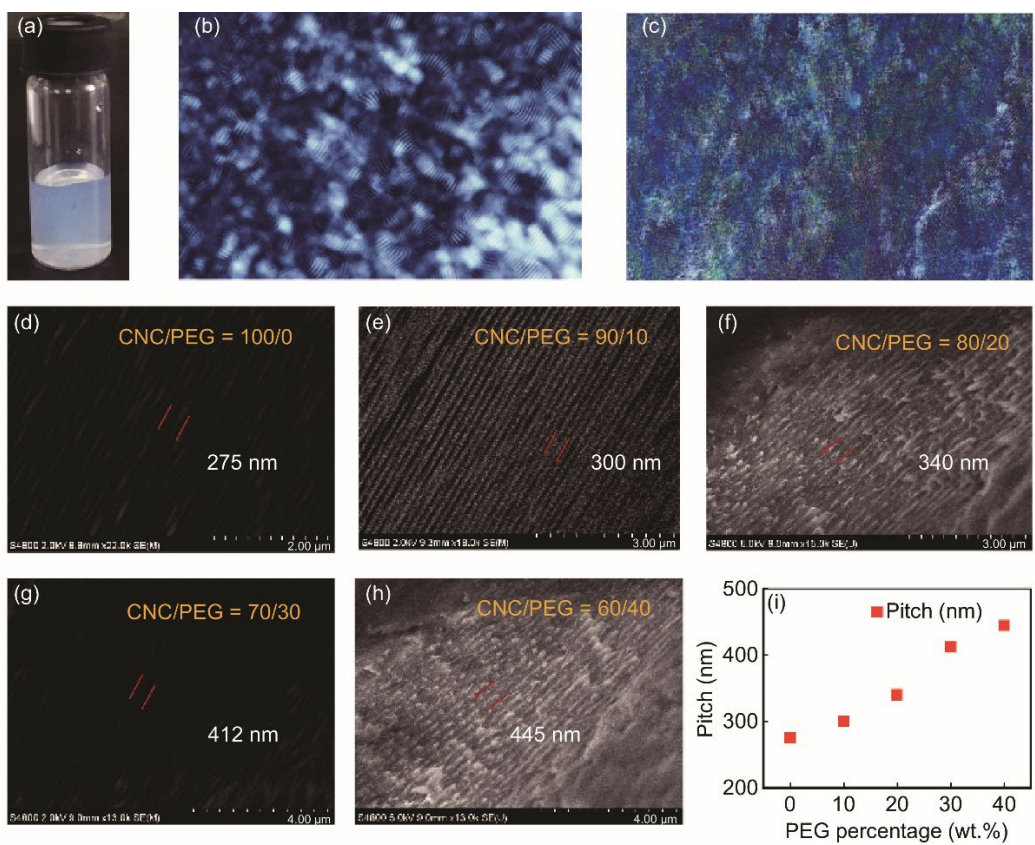

Figure 13. (a) Optical image and (b) POM image of the CNC/PEG/EuW$_{10}$ solution with an 80/20 mass ratio of CNC to PEG, and (c) the corresponding dry film. SEM images of the composite films with different CNC/PEG mass fractions: (d) 100/0, (e) 90/10, (f) 80/20, (g) 70/30, and (h) 60/40.

(i) Plot of the helical pitch as a function of PEG mass ratio in CNC/PEG/EuW$_{10}$ composite films. Copyright 2022 by the authors. Published by SciOpen under a Creative Commons CC BY License.[99]

Similarly, as in many previous studies, this research utilized the Bragg equation to calculate both the photonic band gap and helical pitch. The results showed that as the PEG mass ratio increased, both the helical pitch and photonic band gap exhibited a gradual and linear increase (Figure 14g), indicating a redshift in the photonic band gap and the reflection peak from 422 nm to 666 nm (Figure 14f). This shift also led to noticeable changes in the structural colors of the composite, transitioning from pale purple to red at 13% humidity, as shown in Figure 14a-e.

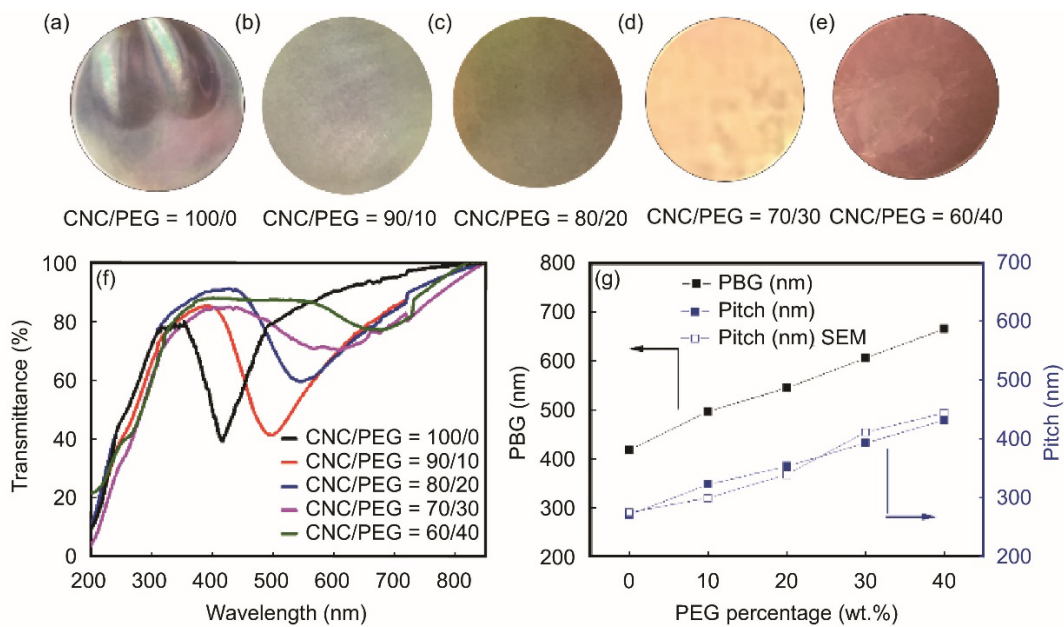

Figure 14. Images of CNC/PEG/EuW$_{10}$ composite films with different CNC/PEG mass fractions at 13% humidity: (a) 100/0, (b) 90/10, (c) 80/20, (d) 70/30, and (e) 60/40. (f) Transmission spectra plot for different CNC/PEG mass fractions. (g) Nearly linear correlation between the photonic band gap, helical pitch, and PEG percentage, as determined using the Bragg equation and SEM

images. Copyright 2022 by the authors. Published by SciOpen under a Creative Commons CC BY License. [99]

The Bragg equation is given:

$$\lambda = n_{av}\, p \sin\theta$$

Here, $\lambda, n_{an}, \theta, and\ p$ represent the photonic band gap, average refractive index, incident angle, and helical pitch, respectively. As shown in Figure 14, there is a good compatibility between the pitch calculated by the Bragg equation and SEM. As shown in Figure 15h, the circular dichroism (CD) spectrum exhibits a peak at higher wavelengths and a trough at lower wavelengths, indicating a positive Cotton effect. All the prepared CNC/PEG/EuW$_{10}$ composite films exhibited strong circularly polarized luminescence (CPL) signals. Specifically, the composite films with CNC/PEG mass ratios of 100/0, 90/10, and 80/20 demonstrated the highest asymmetric factor of -0.37 at the 589 nm emission peak, with a photonic band gap range below 600 nm. This emission is attributed to the magnetic dipole-allowed transition $^5D_0 \rightarrow {}^7F_1$. In contrast, the strongest CPL signal with the highest dissymmetry factor was observed in the composites with CNC/PEG mass ratios of 70/30 and 60/40, occurring at 622 nm, with a photonic band gap range between 600 nm and 680 nm. This peak corresponds to the electric dipole-allowed transition $^5D_0 \rightarrow {}^7F_2$ of EuW$_{10}$, indicating a significant contribution of electric dipole transitions to the optical activity in these composites (Figure 15). According to the formula for the asymmetric factor, when the photonic band gap of CNC overlaps with the fluorescence emission of fluorophores, the forbidden propagation effect amplifies the intensity difference between left- and right-circularly polarized luminescence (L-CPL and R-CPL), leading to a higher asymmetric factor. Consequently, tuning the helical pitch or incorporating suitable dopants can further enhance the asymmetric factor, as confirmed by previous studies, including those by Wang [99] and Zheng[103]. The structural colors of the

CNC/PEG/EuW$_{10}$ composite films exhibited a notable shift as the humidity increased from 13% to 100%, highlighting the hydrophilic nature of the PEG and CNC blend. This humidity-dependent behavior was further confirmed by the transmission spectra, which demonstrated a redshift from 522 nm to 882 nm under 100% relative humidity (RH), with the shift becoming more pronounced as the PEG content increased. Table 3 summarizes all the key findings related to the optical and structural changes observed in the composite films under varying humidity conditions.

Table 3. The comparison of the photonic band gap, the asymmetry factors, and the corresponding emission peaks of the composite films at different humidity and CNC/PEG fraction mass. The data reproduced from Wang et al. [99]

| CNC/PEGs' Mass Fraction of CNC/PEG/EuW$_{10}$ | Asymmetry Factor and its Transitions (RH = 13%) | PBG (nm) / Humidity = 13% | Asymmetry Factor and its Transitions (RH = 100%) | PBG (nm) / Humidity = 100% |
|---|---|---|---|---|
| 100/0 | −0.03 ($^5D_0\rightarrow{}^7F_1$) | 422 | −0.06 ($^5D_0\rightarrow{}^7F_2$) | 522 |
| 90/10 | −0.10 ($^5D_0\rightarrow{}^7F_1$) | 500 | −0.18 ($^5D_0\rightarrow{}^7F_2$) | 631 |
| 80/20 | −0.37 ($^5D_0\rightarrow{}^7F_1$) | 545 | −0.18 ($^5D_0\rightarrow{}^7F_4$) | 686 |
| 70/30 | −0.19 ($^5D_0\rightarrow{}^7F_1$) | 606 | −0.02 ($^5D_0\rightarrow{}^7F_4$) | 720 |
| 60/40 | −0.21 ($^5D_0\rightarrow{}^7F_2$) | 666 | Weak Signal | 882 |

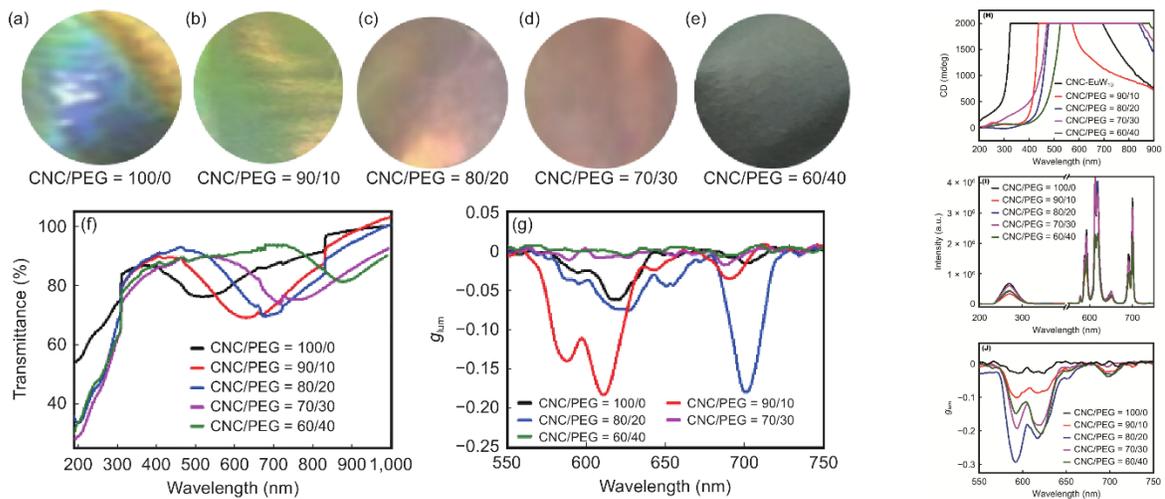

Figure 15. Images of CNC/PEG/EuW$_{10}$ composite films with varying CNC/PEG mass fractions under 100% humidity conditions: (a) 100/0, (b) 90/10, (c) 80/20, (d) 70/30, and (e) 60/40. (f) Transmission spectra and (g) CPL spectra of the composite films, along with their transitions at 100% humidity. (h) Circular dichroism spectra, (i) fluorescence spectra (excitation wavelength = 270 nm), and (j) Circular polarized luminescence spectra of CNC/PEG/EuW$_{10}$ composite films with different CNC/PEG mass fractions at a relative humidity of 13%. Copyright 2022 by the authors. Published by SciOpen under a Creative Commons CC BY License. [99]

In a separate study, Matteo Cei et al.[104] designed two different configurations to investigate the effect of enantiomers and CNC on circularly polarized luminescence (CPL). The configurations were as follows: (i) E-CNC-D Configuration: The emissive layer (E) was placed first, followed by the CNC layer, and finally the detector (D). (ii) CNC-E-D Configuration: The order was reversed, placing the CNC layer first, followed by the emissive layer, and then the detector using spin-coating on a dry CNC film. As one might expect, in the E-CNC-D configuration, the CNC layer may act as a selective filter, reflecting left-handed circularly polarized luminescence while transmitting right-handed CPL. This will result in positive circular extinction, measured as the difference between the left and right circularly polarized optical densities, and a negative dissymmetry factor. Conversely, in the CNC-E-D configuration, the outcome may vary: it either will exhibit a reduced dissymmetry factor and circular extinction, or it will display negative circular extinction with a positive dissymmetry factor. This difference is attributed to the altered sequence in which light interacts with the chiral CNC structure before reaching the emissive layer. They first employed an optically inactive europium complex, $Eu(TTA)_3Phen$ (TTA = thenoyltrifluoroacetone, Phen = 1,10-phenanthroline) in combination with CNC. As expected, in the E-CNC-D configuration, a negative dissymmetry factor of -0.18 was observed, while in the CNC-E-D configuration, a positive dissymmetry factor of 0.19 was recorded. To further investigate the effect of different enantiomers (R,R and S,S) of europium complexes, the researchers examined $Eu(TTA)_3iPrPyBox$ (iPrPyBox = 2,2′-(2,6-pyridinediyl)bis(4-isopropyl-2-oxazoline)). These two enantiomeric complexes exhibited two distinct CPL bands with opposite signs and different dissymmetry factors, specifically $|g_{lum}|$ = 0.02 at 614 nm and 0.2 at 595 nm. In this case, a more complex behavior was expected due to the interplay between the intrinsic circularly polarized luminescence (CPL) of the molecular complex and the structural chirality of CNC. When the intrinsic CPL of the optically active complex aligns with the polarization induced by CNC, an enhanced CPL signal is observed. Conversely, when their effects oppose each other, the CPL intensity decreases. In the CNC-E-D configuration, the (R,R) enantiomer exhibited a weak dissymmetry factor of -0.01 at 595 nm, primarily due to the opposing effects of the left-handed reflective behavior of CNC and the intrinsic polarization of the europium complex. This interference weakened the overall CPL signal. However, in the E-CNC-D configuration, the dissymmetry factors were -0.17 and -0.15 for the (R,R) and (S,S) enantiomers, respectively, at 595

nm—a transition attributed to the magnetic dipole-allowed process. Furthermore, at 614 nm, a negative dissymmetry peak was observed for both enantiomers, corresponding to the electric dipole-allowed transition, with a dissymmetry factor of -0.17. In this scenario, since europium complexes inherently exhibit weak CPL signals ($|g_{lum}| \approx 0.02$), the majority of the observed polarization ($\Delta OD/OD \approx 0.52$) in the system originated from the chiral environment of CNC, rather than from the europium complex itself.[104]

In a separate study, Sun et al.[105] developed a Janus-structured, switchable circularly polarized light (CPL) emitter. They synthesized a blue-emitting fluorescent film by incorporating carbon dots into a polyvinyl alcohol (PVA) matrix. To achieve CPL properties, the fluorophores required a chiral environment, which they introduced by employing cellulose nanocrystals (CNCs). To fabricate the switchable CPL emitter, they coated one side of the fluorescent film with CNCs, a left-handed chiral nematic material that preferentially transmits right-handed CPL. The opposite side was coated with hydroxypropyl cellulose (HPC), a right-handed cholesteric liquid crystal with lyotropic liquid crystal property[106, 107] that selectively transmits left-handed CPL. This Janus structure enabled controllable CPL emission, making it a potential chiral light source for applications such as asymmetric photopolymerization reactions[108] and optical information encryption[109]. This Janus setup resulted in strong left- and right-circularly polarized luminescence (L-CPL and R-CPL) signals, with dissymmetry factor values of 0.28 and -0.68, respectively. [105]

Shi et al.[110] successfully synthesized a thermo-responsive carbon quantum dot (CQD)/hydroxypropyl cellulose (HPC) composite through a self-assembly process at room temperature. The optical characterization of the composite revealed red shifts in the reflection spectra with increasing temperature, as depicted in Figure 16. Since HPC exhibits a right-handed cholesteric nematic phase, the CQD/HPC composite displayed a negative circular dichroism (CD) response. Interestingly, while the circular dichroism (CD) intensity showed a gradual increase beyond −25,000 mdeg, indicating that the composite maintained a well-ordered cholesteric liquid crystal structure up to a certain threshold, a subsequent decline was observed as temperature continued to rise. This behavior suggests that excessive heating disrupts the structural ordering of the cholesteric phase. Moreover, within the temperature range of 20–26°C, a strong overlap between the photonic band gap and the circularly polarized luminescence (CPL) spectrum was observed, leading to enhanced CPL signals. However, as the photonic band gap redshifted further, the decreasing spectral overlap resulted in a decline in CPL intensity. Although both CD and CPL signals initially increased with temperature, the dissymmetry factor exhibited a consistent decline. This trend was attributed to the broadening of the full width at half maximum (FWHM) of the CD, CPL, and dissymmetry factor spectra, which suggests a loss of cholesteric liquid crystal order and an increase in the population of thermal phonons. Under 360 nm UV excitation, the dissymmetry factor decreased from 1.02 at 22°C to 0.8 at 44°C, demonstrating the thermal sensitivity of the composite material.[110]

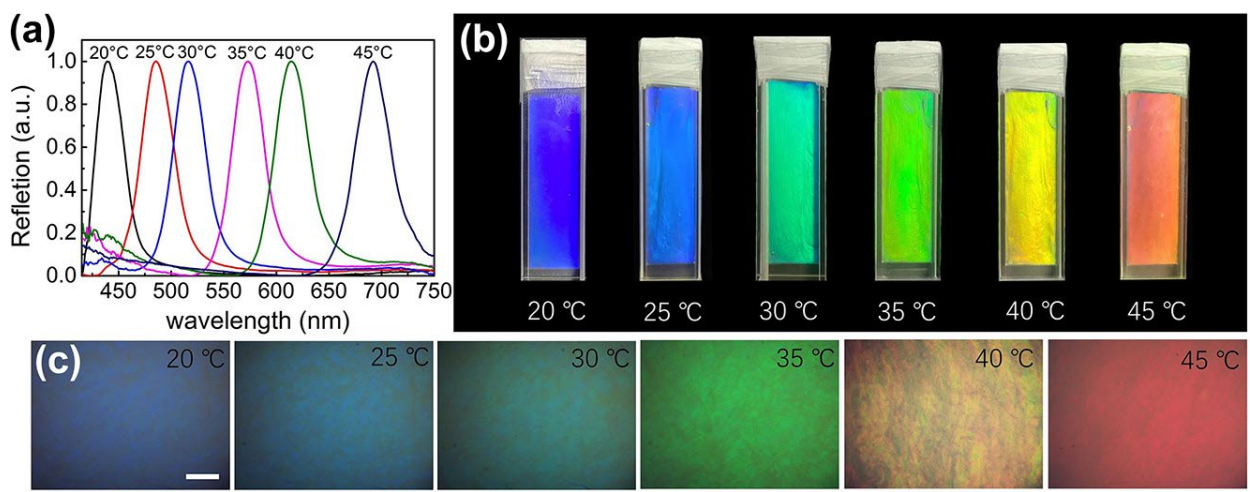

Figure 16. illustrates the temperature-dependent optical properties of the CQD/HPC composite. (a) presents the normalized reflection spectra at varying temperatures, demonstrating a shift in the photonic band gap due to thermally induced structural changes. (b) shows the corresponding color variations observed under natural light. (c) provides reflective microscopic images of the CQD/HPC composite at different temperatures, captured at a 250 μm scale. Copyright 2024 by the authors. Published by De Gruyter under a Creative Commons CC BY License.[110]

In another study, Shaoyi Cao et al.[111] developed a multi-stimuli-responsive CNC/Curcumin composite, where curcumin served as the light-emitting source and CNC functioned as the chiral body. One of the prepared composites (CNC/Cur-0.4) exhibited excitation wavelength-dependent fluorescence spectra, with the highest fluorescence intensity of 160 at 390 nm, which gradually decreased to less than 10 at 490 nm, with a λmax of 550 nm. However, the composite exhibited structural color due to the intrinsic ability of CNC to self-assemble into a cholesteric phase liquid crystal structure, causing the pure CNC film to appear blue. SEM analysis confirmed that the incorporation of CNC and curcumin did not alter the left-handed chiral nematic structure of CNC. The composite displayed blue iridescence under left-handed circularly polarized fluorescence (L-CPF) but showed no iridescence under right-handed circularly polarized fluorescence (R-CPF). In this scenario, sucrose was used to tune the structural colors of the CNC/Cur composite by modifying its iridescence through adjustments in the mass ratio of sucrose. As observed in POM images, the introduction of sucrose shifted the structural colors of the composite from blue to red, indicating an increase in the helical pitch from 263 nm to 495 nm and a red shift in the reflection peak from 335 nm to 750 nm. Additionally, the incorporation of sucrose enhanced the flexibility of the composite. Due to its superb hygroscopic properties, the CNC/Cur composite exhibited humidity-responsive behavior. When the humidity increased from 45% to 98%, the composite experienced a decrease in CPL strength, with the asymmetric factor decreasing from -0.138 to -0.06, along with a red shift in the photonic band gap. This effect can be explained by the overlap between the fluorescence emission band of curcumin and the photonic band gap of the composite. At low humidity levels, this overlap is large, leading to strong CPL. However, as humidity increases, the helical pitch expands, reducing the overlap due to the red shift of the photonic band gap, which weakens the inhibition of left-handed CPL (L-CPL). On the other hand, CNC thin films are highly sensitive to moisture and humidity, which can lead to their degradation. To mitigate this issue, the authors applied a polyvinyl butyral (PVB) coating to enhance the hydrophobicity of the composite. This treatment resulted in a 22° increase in the contact angle, reaching 128° when the PVB coating concentration was increased from 0.5 wt% to 3 wt%. FTIR and XRD analyses confirmed that no structural damage was caused by the PVB coating on the chemical composition of the composite. Additionally, after being immersed in water for up to 10 minutes, the composite showed no significant structural changes, demonstrating the effectiveness of PVB in improving water resistance.[111] However, the pH-responsive behavior of this composite was primarily attributed to the enol–keto tautomeric interconversion in the structure of curcumin under acidic and basic conditions.[112] In another study, Ma et al.[113] synthesized a series of chiral carbon dots (CCDs)/cellulose nanocrystal (CNC) host-guest composite films using a matrix-assisted method. The films exhibited high luminescence dissymmetry (asymmetry) factor values, ranging from -0.24 to -0.90, indicating their potential application in visual polarization detection. Although CNCs emit left-handed circularly polarized luminescence, CCDs, through non-covalent interactions with the soft CNC template, exhibited right-handed CPL along with angle-dependent structural colors. The resulting composite films demonstrated an average fluorescence lifetime ranging from 6.05 ns to 2.46 ns, with photoluminescence quantum yields (PLQY) reaching up to 56.10%. Researchers Sun et al.[114] biosynthesized a series of bacterial cellulose (BC) hybrids exhibiting circularly polarized luminescence (CPL) in the near-infrared, green, red, and orange regions, as shown in Figure 17. A significant amplification of CPL performance was observed, with the dissymmetry factor enhanced by up to $10^{-2}$ scale following in situ bacterial fermentation.

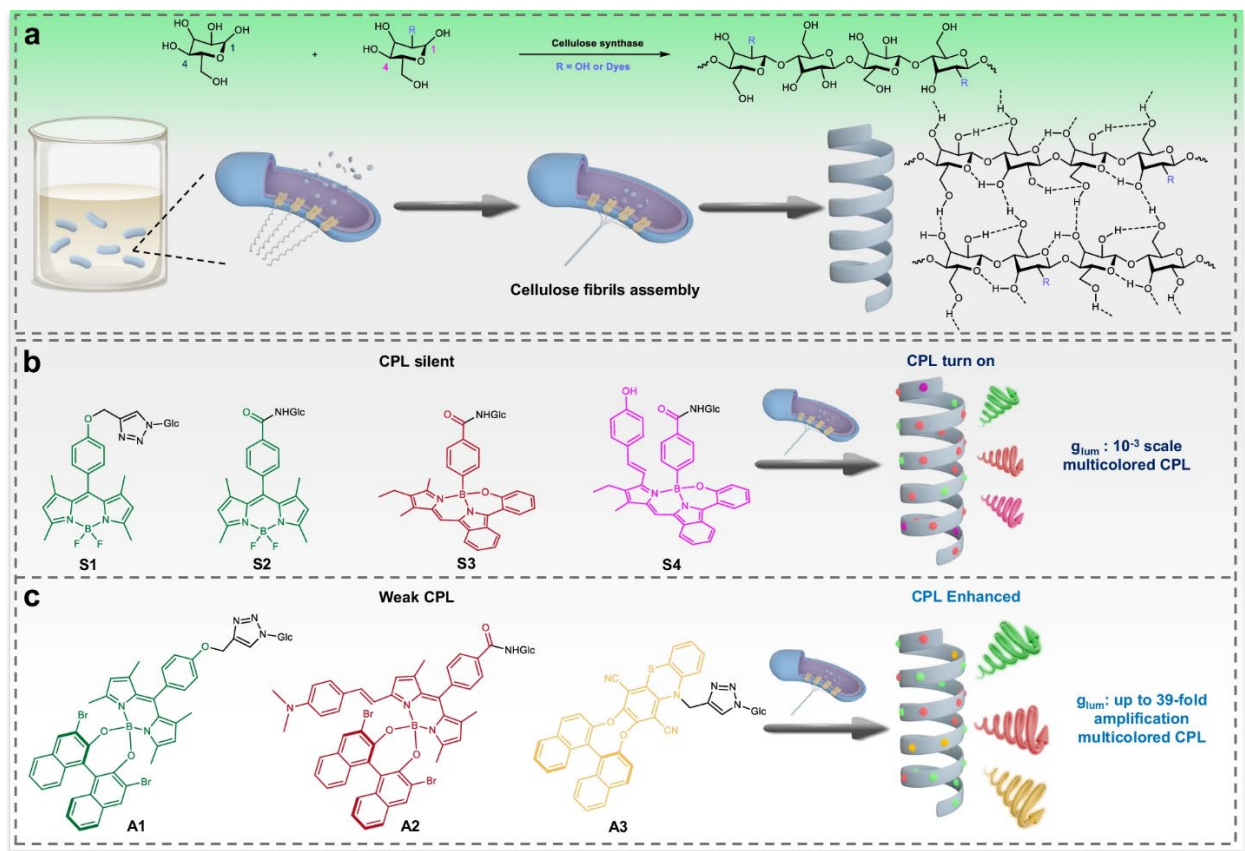

Figure 17. (a) β−1,4-glycosidic reaction catalyzed by cellulose synthase in bacteria. (b) S1, S2, S3, and S4 are CPL-silent glycosylated molecules that exhibit a turn-on response after in situ bacterial fermentation. (c) Weak CPL performance is enhanced after bacterial fermentation, with the dissymmetry factor amplified up to 39-fold. Copyright 2025 the authors. Published by Nature under a Creative Commons CC BY License.[114]

However, the dithienylethene and dansyl chloride combination (DE-DC) serves as an excellent candidate for fluorescence-switching molecules.[115-117] In this system, dithienylethene undergoes cyclization, acting as an energy acceptor to quench the fluorescence of the dansyl group under 254 nm UV light. Due to the ring-opening process, the fluorescence property is restored under 590 nm.[117] Additionally, this blend exhibited minimal toxicity toward bacterial growth, making it a promising candidate for bacterial fermentation applications. Due to these excellent properties, the researchers synthesized the DE-DC hybrid BC biofilm to leverage the light-induced fluorescence changes of DE-DC and proposed information encryption models, as shown in Figure 18. Figure 18b illustrates the photoetching erasure process, where first, upon exposure to 254 nm light, the fluorescence quenching effect creates a "UM" pattern on the biofilm. Then, when the film is exposed to 590 nm light, the "UM" pattern is erased. This process can be repeated, with the term "FHS" being photoetched and erased in a similar manner. The researchers also demonstrated the use of this technique to generate quick response (QR) codes, as shown in Figure 18c. Furthermore, an 8-bit American Standard Code for Information Interchange (ASCII) was successfully encoded using DE-DC Hybrid BC, BC, and S2-BC. Initially, the term "917=" was displayed, but after 254 nm UV light exposure and the fluorescence quenching effect in the DE-DC hybrid BC, the correct information, "1069", was revealed, as shown in Figure 9d.[114]

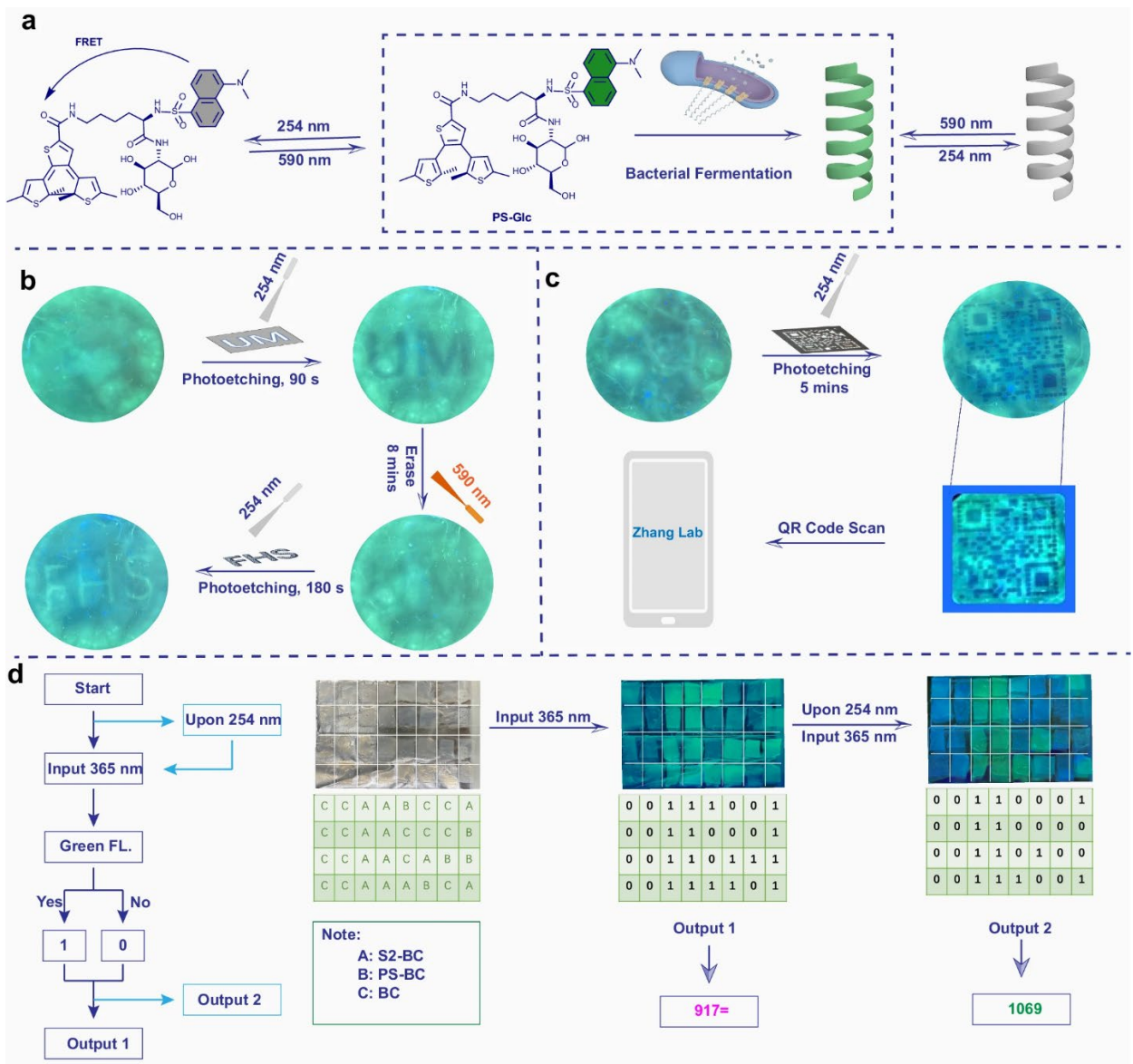

Figure 18. Information encryption applications based on photo-switching bacterial cellulosic biofilms. (a) Photo-switching mechanisms of DE-DC Hybrid BC and DE-DC upon 590 nm and 254 nm irradiation, along with the bacterial-assisted fabrication of DE-DC Hybrid BC. (b) Fluorescence quenching and photoetching process upon 254 nm exposure, followed by the photo-erasure and fluorescence revival process under 590 nm visible LED light. (c) QR code information encryption using the photo-switching mechanism. (d) 8-bit American Standard Code for Information Interchange (ASCII) representation, where blocks with green fluorescence indicate

"1," while fluorescence-quenched regions represent "0." The encoded information is read under a 365 nm hand-held UV lamp, with the flowchart on the left demonstrating the readout protocol. Copyright 2025 the authors. Published by Nature under a Creative Commons CC BY License.[114]

Mascruchin et al. [9] investigated the relationship between the birefringence of CNC, its concentration, solvent type, and ultrasonication time. In a static state without ultrasonication, CNC suspensions in water with concentrations ranging from 0.1% to 8% exhibited no birefringence, indicating that refraction occurred only due to the anisotropic crystal arrangement in the suspension. In a dynamic state (stirred but without ultrasonication), CNC suspensions below 0.1% concentration showed no birefringence, suggesting that at low concentrations, electromagnetic waves transmit almost completely. Birefringence became apparent at concentrations between 0.5% and 2.5%. Upon sonication, all samples except those with less than 0.1% CNC concentration exhibited homogeneous birefringence, indicating that dried agglomerated cellulose dispersed upon defibrillation of nanocrystalline fibrils, leading to improved transparency and birefringence.

The study further examined CNC dispersion in various solvents, including benzene, N,N-dimethylacetamide (DMA), ethanol, dimethylformamide (DMF), acetone, acetic acid (1%), and water. Due to the high crystallinity[118] and hydrophilic surface of CNC, adequate solubility was observed in polar solvents with hydroxyl groups. CNC exhibited partial dispersion in DMA, ethanol, DMF, and acetone, while it was fully diluted in acetic acid (1%) and water. Birefringence was observed only in fully diluted solutions, confirming that solvent interactions and dispersion homogeneity influence the optical properties of CNC suspensions.

Challenges and Future Directions

Nanocellulose has emerged as a promising material for optoelectronic applications due to its renewable sourcing, tunable optical properties, and mechanical robustness. However, several challenges must be addressed to unlock its full potential. A primary limitation lies in its inherent hydrophilicity, which leads to moisture-induced swelling and structural degradation in humid environments. For instance, CNC/PEG composite films exhibit humidity-driven shifts in their photonic bandgap (PBG), destabilizing their optical performance.[99] Additionally, mechanical fragility under thermal or mechanical stress compromises the durability of nanocellulose-based thin films in long-term applications.[110, 119] Scalability remains another hurdle, as energy-intensive production methods such as acid hydrolysis for CNCs and bacterial fermentation for BC increase costs and limit large-scale adoption. However, methods such as acid hydrolysis can result in restrictions on the synthesis of intricate structure with high surface area and control on the dimentionality of nanofibers [120, 121] Achieving uniformity in chiral nematic structures, which is critical for reproducible photonic properties, remains challenging due to variations in helical pitch and alignment during self-assembly. These structures are also sensitive to less controllable factors such as humidity and temperature, particularly in sensor applications.[99, 103, 111] Optical performance limitations further complicate practical implementation. Pure nanocellulose exhibits weak intrinsic nonlinear optical (NLO) responses, such as second harmonic generation (SHG), necessitating doping with plasmonic nanoparticles (e.g., Au, Ag) or quantum dots to enhance these effects.[71, 81, 88, 92] Fluorescence quenching in composites, such as CNC/curcumin systems, further reduces quantum yields due to aggregation or non-radiative decay pathways (Cao et al., 2025).[98, 111] Integration with conventional optoelectronic materials is hindered by refractive index mismatches; most nanocellulose derivatives have low range of indices (1.4–1.6)[21], complicating their use in

Bragg reflectors or waveguides with high-index materials like perovskites (Seike et al., 2024).[22] Poor interfacial compatibility between hydrophilic nanocellulose and hydrophobic substrates (e.g., metals, polymers) also limits device performance.[122] Computational modeling gaps exacerbate these challenges. Traditional optical models, such as the Cauchy and Vuks equations, struggle to predict anisotropic behavior in liquid crystalline or disordered photonic systems.[33] Furthermore, the data quality, data volume, privacy and security of experimental datasets can impedes the training of AI models for predicting complex properties like circularly polarized luminescence (CPL) dissymmetry factors.[123] Despite these challenges, innovative strategies are emerging to advance nanocellulose-based optoelectronics. Surface functionalization via hydrophobic coatings (e.g., polyvinyl butyral) or covalent modifications (e.g., esterification) could mitigate environmental sensitivity while preserving optical clarity (Cao et al., 2025).[111] Bioinspired designs mimicking natural photonic structures—such as the ultra-white scales of the Lepidiota stigma beetle or beetle of Chalcothea smaragdina—could optimize light scattering and structural coloration.[14, 124] Green synthesis methods, including enzymatic hydrolysis and ionic liquid processing, offer eco-friendly alternatives to toxic acids for CNC extraction.[4, 120] Advanced manufacturing techniques like 3D/4D printing could enable programmable chiral photonic crystals by leveraging nanocellulose's shear-thinning behavior.[125] Hybrid material systems integrating plasmonic nanoparticles or 2D materials (e.g., MXenes) may amplify nonlinear effects via localized surface plasmon resonance.[75, 126] AI-driven approaches, such as multi-scale modeling frameworks combining the conventional finite-difference time-domain (FDTD) and machine learning, could accelerate the prediction of anisotropic refractive indices and PBG tuning.[59] Generative AI models, like variational autoencoders (VAEs), might further streamline the design of nanocellulose composites with tailored optical properties.[69] Standardized characterization protocols are critical for progress. Operando techniques, such as in situ Raman spectroscopy and small-angle X-ray scattering (SAXS), could provide real-time insights into structural dynamics under environmental stressors.[50] Emerging applications could exploit nanocellulose's chirality and biocompatibility for quantum photonics (e.g., chiral hosts for single-photon emitters) and transient optoelectronics (e.g., biodegradable solar cells). [109, 114]

In conclusion, nanocellulose's unique properties position it as a transformative material for next-generation optoelectronics. Addressing challenges in stability, scalability, and performance requires interdisciplinary innovation—spanning material engineering, AI-driven design, and sustainable manufacturing. By bridging these gaps, researchers can unlock applications ranging from anti-counterfeiting tags to quantum communication systems, paving the way for a sustainable optoelectronic future.